\documentclass{aastex701}
\usepackage{amsmath}
\usepackage{fvextra}
\usepackage{placeins}
\usepackage{float}

\begin{document}

\title{MARS: A Lightweight Morphology-Aware RFI Segmentation Network for Mask-Guided Mitigation in Radio Astronomy}

\correspondingauthor{Wesley Armour}

\author[0009-0002-6946-7541,gname=Zhaocheng,sname=Gong]{Zhaocheng Gong}
\affiliation{University of Oxford, Department of Engineering Science, Oxford e-Research Centre}
\email{zhaocheng.gong@st-hildas.ox.ac.uk}

\author[0000-0003-2690-6858,gname=Jack,sname=White]{Jack White}
\affiliation{University of Oxford, Department of Engineering Science, Oxford e-Research Centre}
\email{jack.white@eng.ox.ac.uk}

\author[0000-0002-2892-8025,gname=Jayanta,sname=Roy]{Jayanta Roy}
\affiliation{National Centre for Radio Astrophysics (NCRA), Pune 411007, Maharashtra, India}
\email{jroy@ncra.tifr.res.in}

\author[0000-0003-1756-3064,gname=Wesley,sname=Armour]{Wesley Armour}
\affiliation{University of Oxford, Department of Engineering Science, Oxford e-Research Centre}
\email[show]{wes.armour@oerc.ox.ac.uk}


\begin{abstract}
Next-generation radio telescopes generate filterbank data at rates that make storing all observations for later offline mitigation impractical. Mitigation must therefore operate in real or near-real time within the search pipeline while preserving dispersed astrophysical signals. CPU tools fit GPU-centred search pipelines poorly, while neural alternatives can be computationally heavy.

We present MARS, a GPU-based RFI mitigation pipeline centred on a lightweight Morphology-Aware RFI Segmentation Network. The model is a reduced-width, full-resolution U-Net with a bottleneck containing local, horizontal, and vertical filters to capture compact and elongated RFI structures in the frequency-time plane. Normalisation, patch construction, mask reconstruction, replacement, baseline removal, and output rescaling are also implemented on GPU. Training includes an astronomical-signal preservation loss that discourages false flagging of dispersed pulses.

In controlled patch-level tests, MARS achieves an RFI-mask F1 score of \(0.978\) and a precision of \(0.995\). It retains \(97.6\%\) of the injected dispersed-signal fluence in clean patches and \(96.4\%\) of the non-overlapping signal fluence in patches containing mixed injected RFI. Ablation experiments show that the astronomical-signal preservation loss particularly improves the protection of compact, low-DM, high-S/N pulses. At filterbank level, period-matched PRESTO candidates recovered after MARS mitigation have median significance ratios of \(0.90\)--\(0.99\) relative to \texttt{filtool}. Both methods also recover the known pulsars in two real GMRT observations. On an NVIDIA GH200 GPU, MARS achieves a compute-only speedup of \(6.2\times\)--\(7.0\times\) over the fastest tested multi-threaded \texttt{filtool} configurations on an AMD EPYC 9825 CPU.
\end{abstract}



\section{Introduction}
\label{sec:introduction}

Radio frequency interference (RFI) is a major source of contamination in data gathered by radio telescopes and has motivated a long history of mitigation methods for both interferometric and time-domain data \citep{fridman2001rfi,offringa2010aoflagger,offringa2010post}. In pulsar and transient searches, unmitigated RFI can appear as false candidates in acceleration-search results. This can greatly increase the number of candidates that must be inspected and can make weak astrophysical signals harder to identify. RFI mitigation is especially important in pulsar and FRB (Fast Radio Bursts) searches: weak pulsar emission may only be recovered after dedispersion and frequency-domain periodicity or acceleration searches \citep{lorimer2011sigproc,ransom2002fourier}, while FRBs are usually bright millisecond dispersed bursts whose recovered S/N can be reduced if pre-dedispersion mitigation masks part of the pulse track \citep{petroff2019fast,cordes2003searches}. Unmitigated RFI can also generate many false candidates and reduce search sensitivity \citep{morello2022iqrm}. Common RFI morphologies in filterbank data include persistent or intermittent narrowband emission, impulsive narrow or broadband bursts, periodic interference, and mixed time-frequency structures with partial occupancy; in practice, these can also appear as localised patches or fragmented structures in the frequency-time plane \citep{fridman2001rfi,offringa2012morphological,petroff2019fast,yuan2022categorize}. An effective mitigation method must therefore suppress contaminated regions while avoiding unnecessary removal of scientifically meaningful signal.

Computational processing power is also a practical constraint. Modern pulsar and transient pipelines increasingly rely on GPU acceleration for high-throughput processing of large filterbank observations. However, widely used mitigation tools in pulsar-search workflows, including \texttt{filtool}, are primarily CPU-based \citep{men2023pulsarx}. When a CPU mitigation stage is inserted into an otherwise GPU-resident workflow, the mitigation computation itself can become a bottleneck, and additional CPU-to-GPU transfers can reduce the benefit of GPU acceleration in the downstream search. A practical RFI mitigation method should therefore integrate naturally with GPU-centred search pipelines, such as AstroAccelerate \citep{carels2019development}, rather than requiring repeated CPU-GPU data movement. At the same time, producing a standard cleaned filterbank remains useful for standalone offline analysis with established CPU-based tools such as SIGPROC and PRESTO \citep{carels2019development,lorimer2011sigproc,ransom2002fourier}.

Machine-learning methods have shown increasing promise for RFI detection and mitigation, including segmentation models, anomaly-detection approaches, inpainting-based methods, transformer-based architectures, and spiking-neural-network \citep{mesarcik2022learning,van2024remove,du2024comparison,gu2024radio,ouyang2024hierarchical,pritchard2025spiking}. Many of these methods treat RFI detection as an image-like segmentation problem, often drawing on architectures related to U-Net \citep{ronneberger2015u}. These studies demonstrate that neural networks can learn useful RFI representations, but they are not always designed around the deployment constraints of pulsar-search pipelines, where minimal data movement, simple inference operations, and integration with downstream GPU signal-processing modules are central requirements.

We therefore focus on a narrower question: whether the recurring local and anisotropic time-frequency structure seen in many filterbank RFI examples can be captured by a compact deployment-oriented model, rather than requiring a large generic segmentation backbone. Our proposed network predicts a soft RFI mask in the frequency-time plane. It does not directly reconstruct a cleaned observation. The predicted mask is instead applied to the original captured data through an explicit replacement and output-conditioning pipeline, followed by baseline correction and rescaling to produce cleaned data. This separates the learned localisation problem from the replacement policy, making the output easier to compare with classical mitigation tools.

The neural component is a lightweight shape-aware U-Net-style encoder-decoder. The model uses small channel counts, additive skip connections, and standard convolution--batch-normalisation--ReLU (Conv--BN--ReLU) blocks. These operations are deliberately chosen because they are supported by TensorRT \citep{nvidia_tensorrt_docs}, NVIDIA's GPU inference optimisation runtime, which compiles trained neural networks into optimised GPU engines for faster deployment. This allows the trained model to run with low inference overhead inside a GPU-based search pipeline. The bottleneck is RFI shape-aware, using local, horizontal, and vertical convolutional branches to capture common local and anisotropic time-frequency interference structures. The horizontal branches target narrowband structures extended in time, the vertical branches target broadband or short-duration structures extended in frequency, and the local branch targets compact or irregular interference. Additional horizontal and vertical decoder-refinement blocks provide a larger axis-aligned receptive field at high resolution while keeping the operation set compatible with TensorRT deployment.

The model is trained using synthetic frequency-time patches with known RFI masks, where each patch is a contiguous \(512 \times 512\) data segment extracted sequentially from the filterbank, comprising 512 adjacent frequency channels and 512 consecutive time samples. This provides pixel-level supervision without manual annotation and allows the training set to include a controlled range of RFI morphologies. Independently of the synthetic data generation, the pipeline applies robust per-channel normalisation when constructing network inputs. This normalisation places observations with different background levels and noise scales onto a more consistent intensity scale before inference.

A further difficulty is that bright astrophysical transients can resemble interference in a frequency-time image. FRBs and bright single pulses can form high-intensity elongated structures, but unlike local interference, they follow the cold-plasma dispersion relation, \(\Delta t \propto {\rm DM}(f_1^{-2}-f_2^{-2})\). To reduce the risk of suppressing such signals, we inject synthetic dispersed pulse-like signals during training and include an astronomical-signal preservation loss that discourages high-confidence RFI predictions on clean injected-pulse pixels. The goal is not to classify FRBs directly, but to prevent the RFI detector from treating clean dispersed pulse power as interference.

The evaluation is organised into three complementary components. At the patch level, we use Remove First, Detect Later (RFDL) as the neural-network baseline because it was reported as the strongest ANN baseline in a recent comparison of AOFlagger \citep{offringa2010aoflagger}, ANN, and SNN RFI-detection methods on the HERA dataset \citep{van2024remove,pritchard2025spiking}. We therefore compare the proposed mask predictor with RFDL using synthetic RFI masks and injected-pulse protection tests. This isolates the learned mask-prediction component. At the filterbank level, we compare the GPU RFI mitigation pipeline with \texttt{filtool} using synthetic binary-pulsar filterbanks and two real GMRT observations. The cleaned filterbank files are processed using matched PRESTO settings \citep{men2023pulsarx,ransom2002fourier}. Finally, a
compute-time benchmark compares the mitigation runtime of the GPU RFI mitigation pipeline with \texttt{filtool}, while RFDL is included as a neural mask-prediction speed baseline. Thus, RFDL is used as the neural mask-prediction baseline, while \texttt{filtool} is used as the operational cleaned filterbank baseline.

\section{Method}
\subsection{Neural-network architecture}
\label{sec:nn_architecture}

The proposed RFI detector is a lightweight shape-aware encoder-decoder network, in which local and anisotropic convolutions provide explicit sensitivity to compact, time-extended, and frequency-extended RFI structures. It predicts a full-resolution RFI mask from a normalised frequency-time patch. Let \(X\in\mathbb{R}^{1\times H\times W}\) denote the input patch, where \(H\) is the frequency axis and \(W\) is the time axis. The network outputs a single-channel logit map \(Z=f_{\theta}(X)\in\mathbb{R}^{1\times H\times W}\). RFI probabilities are obtained as \(P=\sigma(Z)\), and a binary mask is produced downstream by thresholding, optionally followed by hysteresis post-processing. The network therefore performs RFI localisation rather than direct reconstruction of a cleaned observation.

The backbone is a compact U-Net-like encoder-decoder with channel widths \((8,16,32,64)\). Each feature block is implemented using a TensorRT-compatible convolution--batch-normalisation--ReLU (Conv--BN--ReLU) design. The encoder contains three stride-two downsampling stages, which progressively increase the effective spatial context, while the decoder restores the output to the original \(H\times W\) grid. Decoder features are upsampled using nearest-neighbour interpolation, projected with \(1\times1\) convolutions, and fused with the corresponding encoder features by addition. This additive skip fusion avoids the channel expansion associated with concatenation and reduces memory movement. A final \(1\times1\) convolution maps the highest-resolution decoder feature map to a single RFI-logit channel.

To encode the dominant morphology of input time-frequency data RFI, the deepest feature map is processed by a shape-aware bottleneck. This module contains three parallel convolutional branches: a local \(3\times3\) branch for compact or irregular interference, a horizontal \(1\times9\) branch for narrowband structures extended along time, and a vertical \(9\times1\) branch for broadband or short-duration structures extended along frequency. The branch outputs are concatenated, fused by a \(1\times1\) Conv--BN--ReLU block, and added back to the bottleneck input as a residual correction. This gives the model an explicit inductive bias toward local, time-extended, and frequency-extended RFI structures without requiring a large generic segmentation backbone.

The decoder includes anisotropic refinement at each upsampling scale. We denote the three decoder upsampling stages as \(\mathrm{up2}\), \(\mathrm{up1}\), and \(\mathrm{up0}\), ordered from coarser to finer spatial resolution. For an input patch of size \(H\times W\), these stages produce feature maps at \(H/4\times W/4\), \(H/2\times W/2\), and \(H\times W\), respectively.

After each upsampling stage, the feature map is refined sequentially by horizontal and vertical residual blocks. The horizontal refinement block applies a \(1\times31\) convolution followed by a \(3\times3\) convolution, whereas the vertical refinement block applies a \(31\times1\) convolution followed by a \(3\times3\) convolution. These anisotropic residual modules enlarge the axis-aligned receptive field at the recovered decoder resolutions, helping the model capture thin, horizontally extended narrowband ridges and short-duration broadband structures without substantially increasing the channel width. Their contribution is evaluated using independently retrained
ablation models, with the complete results presented in
Appendix~\ref{app:decoder_ablation}.

The final model contains \(270{,}769\) trainable parameters. During inference, the logit map is converted into an RFI probability map using a sigmoid function and then thresholded to produce a binary mask. This mask is not itself a cleaned observation. Instead, it is passed to a separate replacement pipeline, where the predicted contaminated samples are treated according to an explicit mitigation policy before the data are written as time-frequency data. The principal architectural components and output configuration of MARS are summarised in Table~\ref{tab:mars_architecture}.
\begin{table}[t]
\centering
\caption{Summary of the proposed MARS mask predictor.}
\label{tab:mars_architecture}
\begin{tabular}{ll}
\hline
Component & Configuration \\
\hline
Input & \(1\times H\times W\) frequency-time patch \\
Output & \(1\times H\times W\) RFI-logit map \\
Backbone & Compact U-Net encoder--decoder \\
Channel widths & \((8,16,32,64)\) \\
Feature block & Conv--BN--ReLU block \\
Skip fusion & Additive encoder--decoder skips \\
Shape-aware context & \(3\times3\), \(1\times9\), and \(9\times1\) bottleneck branches \\
Decoder refinement & Residual horizontal and vertical anisotropic blocks \\
Trainable parameters & \(270{,}769\) \\
Inference output & Thresholded RFI mask, with optional hysteresis refinement \\
Mitigation use & Mask-guided replacement of contaminated samples in the original science data \\
\hline
\end{tabular}
\end{table}

\subsection{Preprocessing and patch inference}
\label{sec:preprocessing_patch_inference}

The neural network is applied to fixed-size frequency-time patches rather than to the full filterbank observation at once. This keeps the inference shape fixed, limits GPU memory use, and allows long observations to be processed in temporal segments. Let the raw filterbank data be
\[
R\in\mathbb{R}^{C\times T},
\]
where \(C\) is the number of frequency channels and \(T\) is the number of time samples. For segment \(s\), we denote the corresponding data block by \(R^{(s)}\in\mathbb{R}^{C\times L_s}\), where \(L_s\) is the number of time samples processed in that segment.

In the default configuration, the target segment duration is \(T_{\rm seg}=2\,{\rm s}\). The neural network uses fixed $N_p\times N_p$ patches with
$N_p=512$, corresponding to 512 frequency channels and 512 time
samples per patch. For a given sampling time $t_{\rm samp}$, the
number of time patches per segment is chosen as
\begin{equation}
K=
\max\left(
1,
\operatorname{round}
\left[
\frac{T_{\rm seg}}
     {N_p t_{\rm samp}}
\right]
\right),
\end{equation}
so that the actual segment length is
\begin{equation}
L=K N_p.
\end{equation}
Thus, \(L_s=L\) for each complete segment, whereas the final segment may use its actual processed length. This guarantees that each full segment can be tiled by an integer number of 512-sample time patches, while keeping the segment duration close to the target value.

Before neural-network inference, each segment is normalised per frequency channel. This reduces channel-to-channel gain variations and background offsets, so that the network is driven primarily by local frequency-time morphology rather than by absolute intensity scale. For channel \(c\) in segment \(s\), we compute the median
\[
m_{c,s}=\operatorname{median}_{t}\left(R^{(s)}_{c,t}\right)
\]
and the scaled median absolute deviation
\[
d_{c,s}
=
\frac{\operatorname{median}_{t}
\left|R^{(s)}_{c,t}-m_{c,s}\right|}
{\Phi^{-1}(0.75)}
\approx
1.4826\,\operatorname{median}_{t}
\left|R^{(s)}_{c,t}-m_{c,s}\right|,
\]
where \(\Phi\) is the cumulative distribution function of the standard normal distribution. The factor \(1/\Phi^{-1}(0.75)\) makes the scaled median absolute deviation asymptotically consistent with the standard deviation under Gaussian noise. If \(d_{c,s}\) is below a minimum valid scale \(d_{\min}\), the standard deviation \(\sigma_{c,s}\) of the same channel segment is used instead. The selected scale is therefore
\begin{equation}
a_{c,s}=
\begin{cases}
d_{c,s}, & d_{c,s}\ge d_{\min},\\
\sigma_{c,s}, & d_{c,s}<d_{\min},
\end{cases}
\qquad d_{\min}=0.5.
\end{equation}
To avoid division by zero, we define the safeguarded scale as
\begin{equation}
\widetilde{a}_{c,s}=
\begin{cases}
a_{c,s}, & a_{c,s}>\epsilon_{\rm scale},\\
1, & a_{c,s}\le\epsilon_{\rm scale},
\end{cases}
\qquad
\epsilon_{\rm scale}=10^{-6}.
\end{equation}
Values of \(a_{c,s}\) not exceeding \(\epsilon_{\rm scale}\) are
replaced by unity solely to keep the normalisation well defined. The normalised segment is then
\begin{equation}
Z_{c,t}^{(s)}
=
\frac{R_{c,t}^{(s)}-m_{c,s}}
     {\widetilde{a}_{c,s}}.
\end{equation}
The neural-network input is then obtained by bounded tanh compression,
\[
X^{(s)}_{c,t}=\tanh\left(\frac{Z^{(s)}_{c,t}}{\alpha_{\tanh}}\right),
\]
with \(\alpha_{\tanh}=6\). This compression preserves the ordering of moderate positive and negative deviations while limiting the dynamic range of extreme samples.

The pipeline also includes a deterministic pre-normalisation screening step for degenerate, saturated, and persistently contaminated channel segments. A channel segment is classified as degenerate when its sample standard deviation satisfies \(\sigma_{c,s}\le\epsilon_{\rm scale}\), and as saturated when more than \(50\%\) of its samples exceed a robust segment-level bright threshold. Persistent contamination is identified from outlying channel medians, channel spreads, and bright-sample occupancy, together with their coherence across neighbouring frequency channels. Because these features can be partially flattened by per-channel normalisation, the resulting flags are recorded before the neural-network input is formed. They are retained as auxiliary masks, rather than treated as neural-network predictions, and are later combined with the reconstructed NN mask during replacement.

After normalisation, each segment \(X^{(s)}\in\mathbb{R}^{C\times L}\) is divided into \(512\times512\) patches. Non-overlapping 512-sample blocks are used along the time axis, while complete 512-channel blocks are used along the frequency axis. If \(C\) is not a multiple of 512, an overlapping block is extracted from the final 512 channels. For \(C<512\), blocks from different time intervals are stacked to form a \(512\times512\) network input and separated again after inference.

Patch inference is performed in batches. For each patch \(X_i\), the network produces a logit map \(Z_i=f_{\theta}(X_i)\), and the corresponding probability map is
\[
P_i=\sigma(Z_i).
\]
In standard inference mode, the binary patch mask is obtained by thresholding,
\[
M_i=\mathbb{I}[P_i>\tau],
\]
where \(\tau=0.5\), corresponding to the midpoint of the sigmoid output and, equivalently, to a zero-logit decision boundary. This threshold is fixed before evaluation and applied unchanged to all synthetic patch-level RFI, FRB-protection, and real-data evaluations reported in the main text. These evaluations use the directly thresholded mask so that the reported metrics measure the neural detector itself rather than a post-processing step. An optional hysteresis refinement is examined only in Appendix~\ref{app:hysteresis} and is not used for the results reported in the main text.

The predicted mask for each patch is placed into a segment-level mask using the frequency-channel and time ranges recorded during patch extraction. For each complete 512-channel frequency block, the \(K\) predicted \(512\times512\) patch masks are arranged in temporal order, concatenated along the time axis, and written to the corresponding 512 channels of the segment. If an additional overlapping frequency block is present, only the rows corresponding to channels not already covered by the complete blocks are copied into the segment-level mask. This avoids assigning two predictions to the duplicated channels and yields the reconstructed neural-network mask
\[
M^{(s)}_{\rm NN}\in\{0,1\}^{C\times L}.
\]

Only complete 512-sample time blocks are processed by the neural network. Any remaining temporal tail containing fewer than 512 samples receives no neural-network prediction and is instead treated as masked and assigned the neutral replacement value. The neural-network mask is then combined with the auxiliary masks for blanked, unusable, and pre-flagged channel segments before mask-guided replacement.

\subsection{Mask-guided RFI replacement}
\label{sec:rfi_replacement}

The neural network is used to identify RFI-contaminated samples in the frequency-time data. During inference, it produces an RFI mask, which is subsequently used by the replacement stage to modify the corresponding samples in the filterbank data. The network itself does not synthesise a cleaned observation.

The pipeline uses different normalisation schemes for mask prediction and replacement because the two steps have different objectives. For neural-network inference, each channel segment is normalised using its median and scaled median absolute deviation, and the resulting values are compressed using a bounded tanh transform. This representation reduces channel-to-channel background variations while preserving the local contrast and morphology of RFI structures, making it easier for the network to identify contaminated regions. In contrast, the replacement stage uses a mean/std-normalised data stream derived from the original filterbank data. This choice follows the Filtool-style equalisation used in the comparison pipeline, so that differences in mitigation performance are driven primarily by the predicted mask rather than by a different output scaling convention.

For channel \(c\) in segment \(s\), let
\[
\mu_{c,s}
=
\mathrm{mean}_{t}\left(R^{(s)}_{c,t}\right),
\qquad
\sigma_{c,s}
=
\mathrm{std}_{t}\left(R^{(s)}_{c,t}\right).
\]
For channel segments with \(\sigma_{c,s}>10^{-6}\), the replacement
representation is defined as
\[
S^{(s)}_{c,t}
=
\frac{R^{(s)}_{c,t}-\mu_{c,s}}{\sigma_{c,s}}.
\]
Channel segments with \(\sigma_{c,s}\leq10^{-6}\) are classified as degenerate. Their normalised values are set to zero, and the entire channel segment is retained in the auxiliary mask for subsequent replacement.

For each segment, the reconstructed NN mask is combined with non-neural masks from the preprocessing stage, including blanked, unusable, saturated, and pre-flagged channel segments. Channels with high NN-mask occupancy can also be promoted to segment-level flags. With occupancy threshold \(\rho\), this promotion is defined as
\[
q^{(s)}_c
=
\mathbf{I}
\left[
\frac{1}{L_{\mathrm{seg}}}
\sum_t M^{(s)}_{\mathrm{NN},c,t}
>
\rho
\right],
\]
where \(\rho=0.5\) in the default configuration. The final replacement mask is
\[
M^{(s)}_{c,t}
=
M^{(s)}_{\mathrm{NN},c,t}
\lor
B^{(s)}_{c,t}
\lor
q^{(s)}_c,
\]
where \(B^{(s)}_{c,t}\) denotes the mask generated during preprocessing. If a channel segment is pre-flagged or satisfies \(q^{(s)}_c=1\), all of its time samples are included in the final replacement mask.

In the default configuration, all samples selected by the final mask are replaced by zero in the mean/std-normalised replacement stream:
\[
\tilde{S}^{(s)}_{c,t}
=
\left\{
\begin{array}{ll}
0, & M^{(s)}_{c,t}=1,\\
S^{(s)}_{c,t}, & M^{(s)}_{c,t}=0 .
\end{array}
\right.
\]
Following replacement, the full data stream undergoes optional post-replacement \texttt{zdot} filtering, baseline removal using a \(1\,{\rm s}\) running-median profile, and block-wise rescaling to an 8-bit filterbank with a target mean of 128 and a target standard deviation of 6. The rescaling is applied to all samples to provide a consistent output scale for downstream PRESTO searches and to enable a direct comparison with \texttt{filtool}, which applies an analogous final rescaling. Samples selected by the final mask are then set to the output mean of 128.

\subsection{Optional hysteresis post-processing}
\label{sec:hysteresis}
After neural-network inference, the default binary mask is obtained by thresholding the sigmoid probability map. For a patch-level probability map \(P=\sigma(Z)\), the pure NN mask is
\[
M_{\rm NN}=\mathbb{I}[P>\tau],
\]
where \(\tau=0.5\) in the default configuration. This pure thresholded mask is used for the synthetic patch-level benchmark, so that the reported precision, recall, and F1 score measure the neural network itself rather than the effect of a mask-completion post-processing step.

For real observations, we optionally apply a hysteresis mask-completion step. The motivation is that real RFI is often less regular than the synthetic training masks: long narrowband ridges may be broken, broadband bursts may have weak edges, and structured interference can contain high-confidence cores surrounded by lower-confidence support pixels. A single global threshold can therefore leave small gaps in an otherwise coherent RFI structure. Hysteresis addresses this by using high-confidence pixels as seeds and allowing the mask to grow only through connected lower-confidence support pixels.

Let \(\tau_{\rm seed}\) be the seed threshold and \(\tau_{\rm support}\leq\tau_{\rm seed}\) be the support threshold. In our implementation, \(\tau_{\rm seed}\) defaults to the standard inference threshold \(\tau\), while \(\tau_{\rm support}\) is a configurable lower threshold. We define
\[
S=\mathbb{I}[P>\tau_{\rm seed}],
\qquad
U=\mathbb{I}[P>\tau_{\rm support}],
\]
where \(S\) contains confident RFI pixels and \(U\) contains lower-confidence candidate support. The hysteresis mask is obtained by growing \(S\) through connected pixels in \(U\), using a small anisotropic neighbourhood in frequency and time. The final mask always includes the original high-confidence seed pixels.

This step is deliberately conservative. It is not used in the synthetic patch-level benchmarks because complete ground-truth masks are available and the objective is to evaluate the raw neural detector. Hysteresis is also disabled in the synthetic full-filterbank comparison and in the real-GMRT
results reported in this work. All main results therefore use the probability map thresholded directly at \(\tau=0.5\). Further implementation details and mask-growth safeguards are provided in Appendix~\ref{app:hysteresis}.
Any diagnostic result obtained with hysteresis is labelled explicitly as HYS-enabled.

\subsection{GPU-accelerated implementation}
\label{sec:gpu_acceleration}

The deployment pipeline is designed to keep the data on the GPU, where the downstream GPU-accelerated discovery pipeline operates, and to avoid repeatedly transferring data back to the CPU. After raw filterbank blocks are copied to GPU memory, the segment-level preprocessing, normalisation, patch extraction, mask reconstruction, mask-guided replacement, optional post-replacement \texttt{zdot}, baseline removal, block rescaling, and output preparation are executed using GPU kernels. Only the final output is returned to the CPU for writing to disk in the required filterbank format.

The neural-network inference stage is accelerated using \texttt{TensorRT}. After training, the PyTorch checkpoint is exported to ONNX and compiled into a \texttt{TensorRT} engine with fixed \(512\times512\) spatial input dimensions, matching the patch size used throughout the pipeline. In the production configuration, the engine uses \texttt{FP16} precision with \texttt{FP16} input and output tensors and supports batched inference over multiple patches.

\section{Training Strategy}
\label{sec:training_strategy}

The neural network is trained as a supervised binary segmentation model on synthetic frequency-time patches with known RFI masks. Each training sample consists of an input patch \(X_i\in\mathbb{R}^{1\times512\times512}\), a binary RFI target mask \(Y_i\in\{0,1\}^{1\times512\times512}\), a pixel-weight map \(W_i\), and an optional astronomical-signal support mask \(A_i\). The model outputs a logit map \(Z_i=f_\theta(X_i)\), and the corresponding RFI probability map is \(P_i=\sigma(Z_i)\).

The base dataset contains 4800 training patches and 1200 validation patches. All patches have spatial size \(512\times512\), matching the inference patch size used by the deployment pipeline. The validation set is not augmented during training, so the validation metrics measure the model performance on a fixed set of held-out patches. The independent evaluation datasets described in Section~\ref{sec:evaluation} are generated separately from the training and validation sets.

Training augmentation refers to stochastic modifications applied to training samples during data loading to increase the diversity of examples observed by the model without storing an enlarged dataset. In this work, augmentation is applied online to the normalised training patches, while the validation data remain unchanged. Synthetic RFI structures with horizontal, vertical, compact, block-like, periodic, and composite morphologies are injected with randomly sampled positions, dimensions, and intensities. A fraction of the samples is left without synthetic RFI to provide RFI-negative examples; these samples may contain either the original background alone or an injected dispersed pulse, helping the model reduce false-positive detections and preserve astrophysical signals.

\begin{figure*}[t]
\centering
\includegraphics[width=0.95\textwidth]
{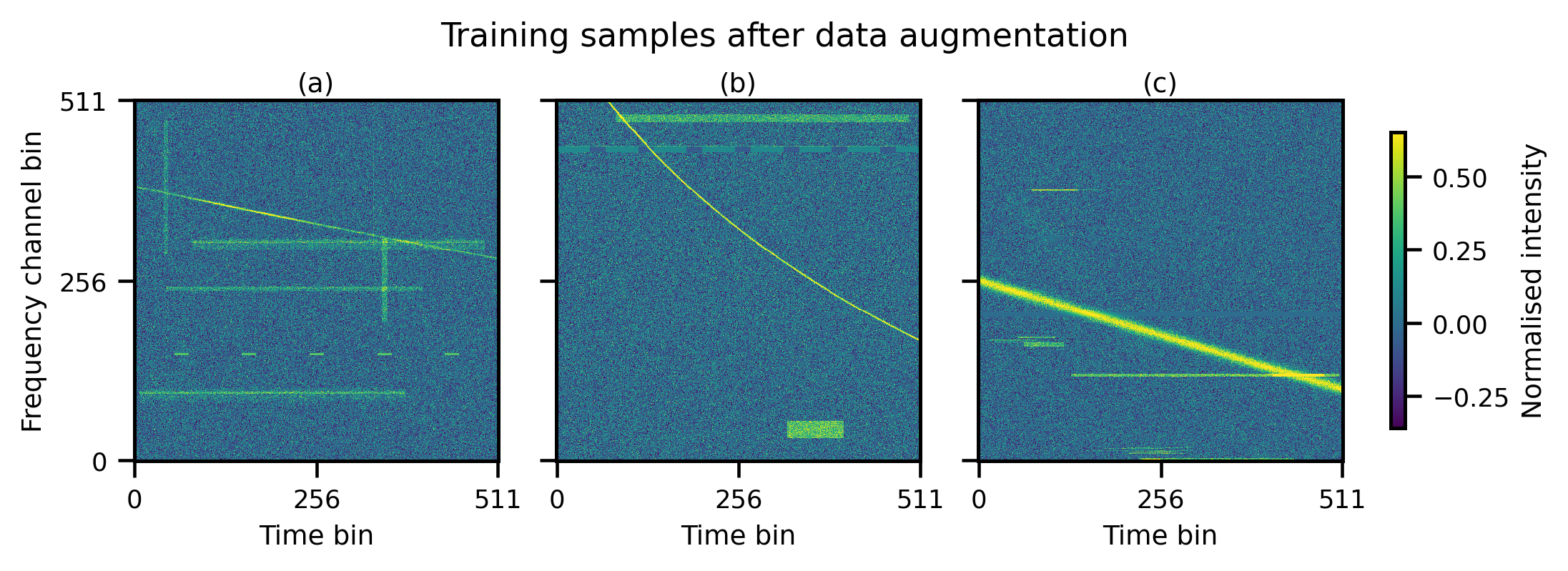}
\caption{Representative training samples after data augmentation. Each \(512\times512\) frequency-time patch was generated from a non-empty training-set background using the same augmentation configuration employed during MARS training. Panel (a) contains a diffuse persistent bright band and vertical bursts; panel (b) contains rectangular blocks and persistent narrowband RFI; and panel (c) contains sparse bright horizontal bars and complex horizontal RFI. The diagonal dispersed tracks are injected astrophysical pulses and are excluded from the RFI target mask. All panels use the same normalised-intensity scale.}
\label{fig:training_augmentation_examples}
\end{figure*}

Synthetic FRB and single-pulse signals are also injected during training. These injections are used to teach the model that bright dispersed pulse power should not automatically be treated as RFI. The FRB augmentation samples over the time-resolution grid used in the evaluation, \(128,256,512,1024\), and \(1310\,\mu{\rm s}\), and uses millisecond-based pulse-width sampling rather than a fixed number of time bins. In the production configuration, the main FRB pulse-width range is \(1\)--\(10\,{\rm ms}\). The augmentation also includes low-DM and bright-pulse branches to target cases that are most easily confused with interference. The astronomical support mask \(A_i\) marks injected pulse pixels that are not part of the RFI target mask. Representative training samples generated by this training augmentation procedure are shown in Figure~\ref{fig:training_augmentation_examples}.

The segmentation loss combines weighted binary cross-entropy, Dice loss, and an astronomical-signal preservation term. Let $B$ be the batch size and $N$ the number of pixels in each patch. The weighted binary cross-entropy is
\[
\mathcal{L}_{\rm BCE}
=
-\frac{1}{BN}
\sum_{b=1}^{B}\sum_{j=1}^{N}
W_{b,j}
\left[
w_+Y_{b,j}\log P_{b,j}
+
(1-Y_{b,j})\log(1-P_{b,j})
\right],
\]
where $b$ indexes patches, $j$ indexes pixels, $P_{b,j}$ is the predicted RFI probability, and $w_+$ is the positive-class weight. The Dice loss is computed for each patch and then averaged over the batch:
\[
\mathcal{L}_{\rm Dice}
=
1-
\frac{1}{B}
\sum_{b=1}^{B}
\frac{
2\sum_{j=1}^{N}W_{b,j}P_{b,j}Y_{b,j}+\epsilon
}{
\max\!\left(
\sum_{j=1}^{N}W_{b,j}(P_{b,j}+Y_{b,j}),
\epsilon
\right)+\epsilon
}.
\]
The astronomical preservation loss penalises high RFI probability on clean injected-pulse pixels:
\[
\mathcal{L}_{\rm astro}
=
\frac{
\sum_{b=1}^{B}\sum_{j=1}^{N}
P_{b,j}^{2}A_{b,j}(1-Y_{b,j})
}{
\max\!\left(
\sum_{b=1}^{B}\sum_{j=1}^{N}
A_{b,j}(1-Y_{b,j}),
\epsilon
\right)
}.
\]
If a batch contains no astronomical support pixels, the same penalty is applied over non-RFI pixels so that the term remains well defined.

The total training objective is
\[
\mathcal{L}
=
\mathcal{L}_{\rm BCE}
+
\lambda_{\rm Dice}\mathcal{L}_{\rm Dice}
+
\lambda_{\rm astro}\mathcal{L}_{\rm astro}.
\]
The model uses \(\lambda_{\rm Dice}=1.0\) and \(\lambda_{\rm astro}=1.0\). To quantify the contribution of the astronomical-signal preservation term, we conducted a controlled loss ablation by retraining the model with $\lambda_{\rm astro}=0$. The network architecture, training and validation data, augmentation settings, random seed, optimiser, learning rate, batch size, and training schedule were kept unchanged. Consequently, the performance difference between the two models isolates the effect of $\mathcal{L}_{\rm astro}$.

\begin{figure}[t]
\centering
\includegraphics[width=0.95\linewidth]{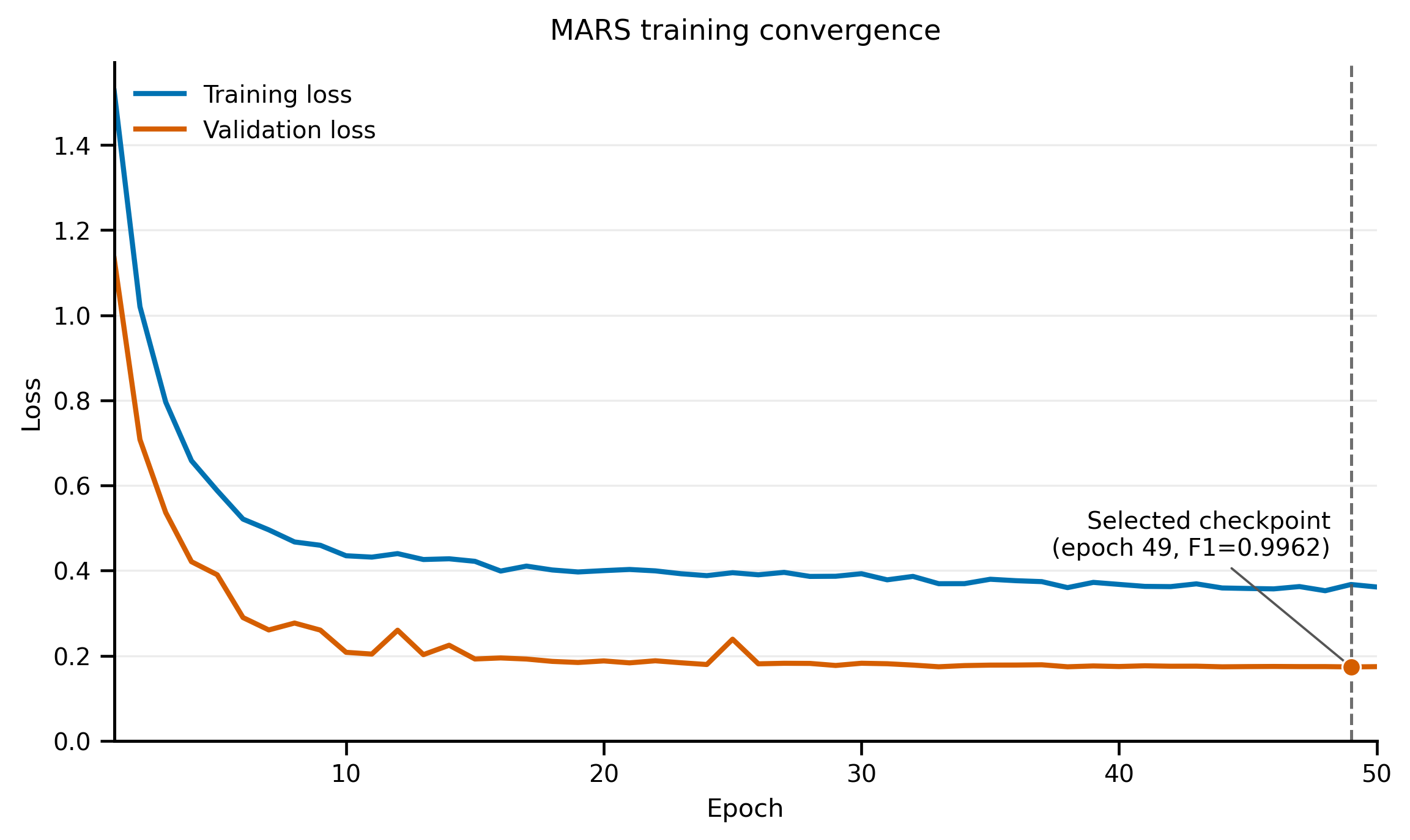}
\caption{Training and validation loss of MARS over 50 epochs. Both losses decrease rapidly during the initial epochs and subsequently stabilise, indicating convergence within the adopted training budget. The dashed vertical line marks the checkpoint selected for the main experiments, corresponding to the highest validation F1 score (\(0.9962\)) at epoch 49.}
\label{fig:mars_training_convergence}
\end{figure}

The model was trained for 50 epochs using AdamW with an initial learning
rate of \(10^{-3}\), weight decay \(10^{-4}\), and batch size 64. The initial learning rate was chosen as a commonly used starting value for 
AdamW and was retained because it produced a stable decrease in both training and validation loss without signs of optimisation instability. A positive-class weight of \(w_+=3.0\) was used to account for the imbalance between RFI and background pixels. When the validation F1 score plateaued for four epochs, the learning rate was reduced by a factor of 0.5 to a minimum of \(10^{-6}\). This validation-driven schedule avoided prescribing a fixed decay point. As shown in Figure ~\ref{fig:mars_training_convergence}, the training and validation losses stabilised within the 50-epoch training budget, with only marginal changes in validation F1 during the final epochs. All 50 epochs were completed, and the checkpoint with the highest validation F1 was selected for the main experiments.

At the end of each epoch, we evaluate the model on the fixed validation set and record loss, precision, recall and F1. Hysteresis post-processing is not used during training or validation; all validation metrics are computed from the raw thresholded neural-network mask.

\section{Evaluation Protocol}
\label{sec:evaluation}

The evaluation is organised at three levels. First, we evaluate the neural mask predictor itself using fixed-seed patch-level benchmarks with known ground-truth masks. These tests compare MARS with the RFDL neural baseline and do not involve replacement, rescaling, or PRESTO searching. Second, we evaluate the complete mitigation pipeline at the filterbank level by applying GPU RFI mitigation pipeline and \texttt{filtool} to the same input filterbanks, writing cleaned \texttt{.fil} files, and searching them with matched PRESTO settings. Third, we compare the mitigation compute time of GPU RFI mitigation pipeline against the fastest tested multi-threaded \texttt{filtool} configuration. Throughout this section, MARS denotes the neural mask predictor, while GPU RFI mitigation pipeline denotes the complete filterbank-mitigation pipeline that applies the mask, performs replacement, and writes a cleaned filterbank. Unless explicitly stated, all neural-network evaluations use the raw thresholded mask with \(\tau=0.5\), and hysteresis post-processing is disabled.

\subsection{Synthetic RFI mask benchmark}
\label{sec:synthetic_rfi_benchmark}

We first evaluate patch-level RFI segmentation using synthetic RFI with known ground-truth masks. This benchmark tests whether the neural mask predictor can identify a range of RFI morphologies under controlled frequency--time conditions. This evaluation uses SPECTRALib to inject diverse RFI.  Further details of the SPECTRALib injection procedure and ground-truth mask construction are provided in Appendix~\ref{app:spectralib_generation}.

For each trial, a clean \(512\times512\) frequency--time patch is selected as the background. The patch is mapped to a raw filterbank-like intensity representation, synthetic RFI is injected in that raw domain, and the contaminated patch is then converted back to the tanh-compressed input representation used by the neural network. The ground-truth RFI mask is defined from the injected RFI support. This allows direct pixel-level evaluation of the predicted mask without any post-processing.

The benchmark includes six RFI families: persistent narrowband, repeating narrowband, impulsive narrowband, repeating broadband, impulsive broadband, and mixed RFI. The mixed-RFI class contains multiple RFI components within the same patch. To test robustness across observing cadence, the benchmark is repeated at five sampling times:
\[
128,\;256,\;512,\;1024,\;{\rm and}\;1310\,\mu{\rm s}.
\]
The complete synthetic RFI benchmark configuration is summarised in Table~\ref{tab:synthetic_rfi_benchmark}.
For each RFI family and sampling-time combination, 500 fixed-seed patches are generated, giving
\[
6\times5\times500=15000
\]
synthetic RFI test patches in total. Repeating and impulsive RFI durations are specified in physical time and converted to sample bins for each sampling time, so that the apparent frequency--time morphology changes consistently with the observing condition.

For each model, the predicted probability map is thresholded at \(\tau=0.5\) to produce a binary RFI mask. We report patch-level precision, recall, and F1 score:
\[
{\rm Precision}=\frac{TP}{TP+FP},\qquad {\rm Recall}=\frac{TP}{TP+FN},
\]
\[
{\rm F1}=\frac{2\,{\rm Precision}\,{\rm Recall}}{{\rm Precision}+{\rm Recall}},\qquad
\]
Hysteresis post-processing is disabled in this benchmark so that the metrics measure the raw neural detector rather than a mask-completion step.

We compare MARS with an RFDL baseline on the same SPECTRALib test patches. RFDL is used here as a pure neural mask-prediction baseline, following previous neural RFI comparisons that identify it as a strong RFI-detection model \citep{van2024remove,pritchard2025spiking}. To make the comparison controlled, RFDL is trained using the same training patches and the same data-augmentation configuration as MARS. Both models are then evaluated on the identical fixed-seed SPECTRALib candidate set using the same threshold and metrics. This comparison therefore reflects differences in the learned mask predictor rather than differences in training data or augmentation policy. The controlled RFDL training configuration and the model-specific preprocessing retained for each method are detailed in Appendix~\ref{app:rfdl_training_config}.

\begin{table}[t]
\centering
\caption{Synthetic RFI mask benchmark.}
\label{tab:synthetic_rfi_benchmark}
\begin{tabular}{ll}
\hline
Item & Setting \\
\hline
Patch size & \(512\times512\) \\
RFI families & 6 families, including mixed RFI \\
Sampling times & 128, 256, 512, 1024, \(1310\,\mu{\rm s}\) \\
Patches & 15000 total; 500 per family/time setting \\
Compared models & MARS, RFDL \\
Metrics & Precision, recall, F1 \\
\hline
\end{tabular}
\end{table}

\subsection{Astrophysical-signal protection benchmark}
\label{sec:astro_signal_protection}

We next evaluate whether the RFI mask predictor preserves injected dispersed astrophysical pulse signals. This is a patch-level benchmark. Synthetic dispersed FRB or single-pulse signals are injected into clean \(512\times512\) frequency--time patches, and the predicted RFI mask is used to measure how much of the injected pulse is retained. The purpose of this test is different from the synthetic RFI mask benchmark: here the injected dispersed signal is not RFI and should therefore remain unmasked.

We consider two test cases. In the first case, the patch contains only the injected dispersed signal on a clean background. This measures the intrinsic tendency of the model to confuse a clean astrophysical signal with RFI. In the second case, both an injected dispersed signal and mixed RFI are present in the same patch. This tests whether the model can preserve the
dispersed signal when complex RFI structures are also present. Partial overlap between the injected signal and RFI is permitted, but is limited to at most \(15\%\) of the injected-signal support. This prevents most of the pulse from being physically contaminated by RFI, which would make the signal-protection
measurement uninformative.

The injected signal grid spans multiple observing conditions. We use five sampling times,
\[
128,\;256,\;512,\;1024,\;{\rm and}\;1310\,\mu{\rm s},
\]
thirteen dispersion measures,
\[
10,\;20,\;30,\;50,\;75,\;100,\;150,\;300,\;600,\;1000,\;1500,\;2000,\;{\rm and}\;3000\,{\rm pc\,cm^{-3}},
\]
We use three nominal peak signal-to-noise ratios,
\((\mathrm{S/N})_{\rm peak}=6, 10,\) and \(20\), defined as the peak
pulse amplitude relative to the robust background noise scale in the
pre-compression \(z\) domain. The three pulse widths are 1.31, 5.24, and
\(15.72\,{\rm ms}\). For each combination of sampling time, DM, S/N, and pulse width, we generate \(N_{\rm rep}=16\) deterministic realisations using fixed seeds. The full astrophysical-signal protection benchmark grid is summarised in Table~\ref{tab:signal_protection_benchmark}. These repeats vary the background patch and pulse placement while keeping the benchmark reproducible.

For the clean-signal benchmark, this gives
\[
5\times13\times3\times3\times16=9360
\]
test patches. For the signal-plus-RFI benchmark, the same signal grid is combined with mixed RFI, giving 9360 signal-plus-mixed-RFI patches. Hysteresis post-processing is disabled in both cases, so the reported signal-protection metrics reflect the raw thresholded neural-network mask.

Let \(M_{{\rm sig},i}\in\{0,1\}\) denote the evaluated injected-signal support mask, let \(m_i\in\{0,1\}\) denote the predicted RFI mask after thresholding at \(\tau=0.5\), and let \(s_i\geq0\) denote the injected pulse fluence contribution at pixel \(i\). For the mixed-RFI benchmark, injected signal pixels that overlap the ground-truth RFI mask are excluded from
\(M_{{\rm sig},i}\), since masking these pixels represents correct RFI detection rather than false suppression of uncontaminated pulse emission. The retained fluence fraction is defined as
\[
R_{\rm fluence}=\frac{\sum_i s_i M_{{\rm sig},i}(1-m_i)}{\sum_i s_i M_{{\rm sig},i}}.
\]
This quantity measures the fraction of injected dispersed-pulse fluence that remains unmasked after applying the predicted RFI mask. Larger values indicate better astrophysical-signal preservation.

The same benchmark is used to evaluate MARS, the RFDL baseline, and the no-\(\mathcal{L}_{\rm astro}\) ablation. The no-\(\mathcal{L}_{\rm astro}\) model uses the same architecture and training setup as the final model but removes the astronomical-signal preservation term from the loss function. This comparison isolates the contribution of \(\mathcal{L}_{\rm astro}\) to signal protection.

\begin{table}[t]
\centering
\caption{Astrophysical-signal protection benchmark.}
\label{tab:signal_protection_benchmark}
\begin{tabular}{ll}
\hline
Item & Setting \\
\hline
Test cases & Clean signal-only; signal plus mixed RFI \\
Sampling times & 128, 256, 512, 1024, \(1310\,\mu{\rm s}\) \\
DM values & 13 values from 10 to \(3000\,{\rm pc\,cm^{-3}}\) \\
S/N values & 6, 10, 20 \\
Pulse widths & 1.31, 5.24, \(15.72\,{\rm ms}\) \\
Repeats & 16 per grid point \\
Patches & 9360 clean-signal; 9360 signal-plus-RFI \\
Compared models & MARS, RFDL, no-\(\mathcal{L}_{\rm astro}\) \\
Metric & Retained fluence fraction \\
\hline
\end{tabular}
\end{table}

\subsection{Synthetic full-filterbank PRESTO comparison}
\label{sec:synthetic_filterbank_presto}

The patch-level benchmarks test mask prediction and signal preservation under controlled conditions, but they do not test whether a cleaned filterbank remains useful for a pulsar-search workflow. We therefore also evaluate the full mitigation pipeline using synthetic SIGPROC filterbank files containing binary pulsar signals \citep{lorimer2011sigproc}. These files are contaminated with SPECTRALib RFI, cleaned independently using \texttt{filtool} and GPU RFI mitigation pipeline, and then searched with PRESTO using matched settings \citep{men2023pulsarx,ransom2002fourier}.

The SIGPROC fake-data generator is used to create binary-pulsar filterbanks at five sampling times: 128, 256, 512, 1024, and \(1310\,\mu{\rm s}\). Each file has 4096 frequency channels, 8-bit samples, a 300 s observation duration, \(f_{\rm ch1}=1550\,{\rm MHz}\), and channel offset \(f_{\rm off}=-0.292968752\,{\rm MHz}\). The pulsar period, pulse width, peak S/N, DM, and binary orbital parameters are drawn from the ranges listed in Appendix~\ref{app:sigproc_fake_config}. The lower bound of the period range is increased for coarser sampling times to avoid generating signals that are poorly represented at the selected cadence.

After the clean binary-pulsar filterbank is generated, SPECTRALib RFI is injected to create a contaminated input file. The same contaminated file is then passed through GPU RFI mitigation pipeline and the \texttt{filtool} baseline. Both methods write cleaned filterbank files. The comparison is therefore performed at the filterbank level rather than at the pixel-mask level.

Both cleaned outputs are searched using matched PRESTO settings. The default synthetic binary-search configuration is \(z_{\max}=200\), \(\texttt{numharm}=8\), \(\texttt{numdms}=1\), \(\texttt{nobary=True}\), \(\texttt{zerodm=False}\), and \(\texttt{rednoise=True}\). The synthetic full-filterbank benchmark and matched PRESTO configuration are summarised in Table~\ref{tab:synthetic_filterbank_benchmark}. The search flow uses \texttt{prepsubband}, followed by \texttt{realfft}, red-noise removal, and \texttt{accelsearch}. Candidate matching is performed using the injected target period or frequency, with a small tolerance and harmonic matching. We therefore compare the recovery of the expected target candidate in each PRESTO output file.

For each sampling time, 200 fixed-seed contaminated filterbanks are generated and processed by both methods. Some injected signals are too weak to be recovered by the PRESTO acceleration search; therefore, the significance comparison includes only realizations in which the injected target candidate is detected in both cleaned outputs. The primary metric is the matched target-candidate significance recovered by PRESTO. We also report the ratio of the GPU-NN-cleaned to \texttt{filtool}-cleaned target significance, which measures whether the GPU RFI mitigation pipeline preserves the injected binary pulsar candidate at a level comparable to the operational baseline.

\begin{table}[t]
\centering
\caption{Synthetic full-filterbank PRESTO comparison.}
\label{tab:synthetic_filterbank_benchmark}
\begin{tabular}{ll}
\hline
Item & Setting \\
\hline
Input & Synthetic SIGPROC binary-pulsar filterbanks \\
RFI injection & SPECTRALib RFI \\
Sampling times & 128, 256, 512, 1024, \(1310\,\mu{\rm s}\) \\
Filterbanks & 200 per sampling time \\
Compared outputs & \texttt{filtool}-cleaned and GPU-NN-cleaned filterbanks \\
PRESTO search & \(z_{\max}=200\), \texttt{numharm}=8, \texttt{numdms}=1 \\
Red-noise removal & Enabled \\
Metric & Matched target-candidate significance \\
\hline
\end{tabular}
\end{table}

\subsection{Real GMRT observations and downstream PRESTO search}
\label{sec:real_gmrt_presto}

Real-observation validation is performed on two GMRT filterbank observations containing known pulsar signals: J0139+5814 and J0659+1414. These observations are used as initial real-data checks of whether the mitigation pipeline preserves known pulsar recovery after realistic RFI cleaning. They are not intended to represent a complete survey of all real RFI environments. For each observation, we generate cleaned filterbanks using GPU RFI mitigation pipeline and the \texttt{filtool} baseline. The resulting filterbanks are then searched with matched PRESTO settings so that differences in recovered candidates are attributable to the mitigation stage rather than to the search configuration.

For J0139+5814, the search is performed at \({\rm DM}=73.81\,{\rm pc\,cm^{-3}}\), with \(z_{\max}=0\) and \(\texttt{numharm}=8\). For J0659+1414, the search is performed at \({\rm DM}=14.05\,{\rm pc\,cm^{-3}}\), with \(z_{\max}=0\) and \(\texttt{numharm}=16\). In both cases, the \texttt{filtool}-cleaned and GPU-NN-cleaned filterbanks are searched using the same PRESTO settings. Since \(z_{\max}=0\), these real-GMRT tests should be interpreted as non-accelerated periodicity searches performed with PRESTO \texttt{accelsearch}, rather than as acceleration searches over a nonzero acceleration range. The source parameters and matched PRESTO settings are summarised in Table~\ref{tab:real_gmrt_settings}.

Each cleaned filterbank is processed with \texttt{prepsubband}, followed by \texttt{realfft}, red-noise removal, and \texttt{accelsearch}. All PRESTO comparisons use matched DM, \(z_{\max}\), harmonic-summing, and red-noise settings for the files being compared. The main real-data metric is the recovered significance of the known target candidate.

\begin{table}[t]
\centering
\caption{Real GMRT observations and PRESTO settings.}
\label{tab:real_gmrt_settings}
\begin{tabular}{lccc}
\hline
Source & DM \(({\rm pc\,cm^{-3}})\) & \(z_{\max}\) & \texttt{numharm} \\
\hline
J0139+5814 & 73.81 & 0 & 8 \\
J0659+1414 & 14.05 & 0 & 16 \\
\hline
\end{tabular}
\end{table}

\subsection{\texttt{filtool} baseline}
\label{sec:filtool_baseline}

The main traditional baseline is \texttt{filtool} from the PulsarX environment \citep{men2023pulsarx}. Since \texttt{filtool} is a full-filterbank mitigation tool rather than a patch-level mask predictor, it is not used in the patch-level SPECTRALib or signal-protection mask benchmarks. Instead, we compare \texttt{filtool} and GPU RFI mitigation pipeline at the cleaned-filterbank level: each method produces a cleaned \texttt{.fil} file, and the resulting files are searched with matched PRESTO settings.

The command template used for the \texttt{filtool} baseline is:
\begin{verbatim}
filtool -t <NTHREADS> -l 2 --baseline 1.0 --zapthre 4.0 \
  --rfi zdot mask 2 2 kadaneF 2 2 kadaneT 2 2 \
  -o <OUTPUT_ROOT> -f <INPUT.fil>
\end{verbatim}
Here \texttt{zdot} is the \texttt{filtool} command option corresponding to the zero-DM matched filter. The same command template is used for all files in the comparison.

GPU RFI mitigation pipeline is configured to use a comparable output convention. Masked samples are zero-filled in the mean/std-normalised replacement stream, optional post-replacement zero-DM matched filtering can be applied, final baseline correction uses a 1.0 s baseline width, and the output is rescaled to an 8-bit filterbank. Masked pixels are forced to the output mean after rescaling, so that replaced samples are written as neutral mean-level pixels in the final \texttt{uint8} output. The comparison with \texttt{filtool} should therefore be interpreted as an operational cleaned-filterbank comparison, not as a pixel-level mask comparison.

\subsection{RFI-mitigation compute-time benchmark}
\label{sec:compute_time_benchmark}

The runtime comparison focuses on the compute time required for RFI mitigation and neural mask prediction. We compare three systems: the proposed GPU RFI mitigation pipeline, the RFDL neural baseline, and \texttt{filtool}. The GPU RFI mitigation pipeline and \texttt{filtool} are operational filterbank-mitigation pipelines that produce cleaned filterbank outputs, whereas RFDL is included as a neural mask-prediction speed baseline. File I/O, model or TensorRT-engine loading, output writing, and downstream PRESTO searching are excluded from the reported timing. This isolates the computational cost of the mitigation or mask-prediction stage under a common compute-only convention. The compute-time benchmark protocol is summarised in Table~\ref{tab:runtime_protocol}.

The GPU RFI mitigation pipeline and the RFDL baseline are evaluated on an NVIDIA GH200 system. The CPU \texttt{filtool} baseline is evaluated on an AMD EPYC 9825 CPU, with multiple thread counts tested to identify the fastest CPU configuration. The comparison therefore includes the TensorRT-accelerated GPU implementation of the proposed pipeline, the GPU runtime of the RFDL neural baseline, and the fastest observed multi-threaded CPU \texttt{filtool} configuration. This is a system-level compute comparison; it is not intended as a hardware-independent algorithmic speedup.

The benchmark file contains 4096 frequency channels and an observation duration of approximately 100 s. For the GPU RFI mitigation pipeline, the reported compute time includes the mitigation stages executed after the input data are available in memory: preprocessing and normalisation, patch construction, NN inference, mask reconstruction, mask-guided replacement, zero-DM matched filtering, baseline correction, and rescaling. For RFDL, the reported compute time includes the preprocessing and neural forward pass needed to generate the RFI mask on the same input data. Since RFDL is used as a neural mask-prediction baseline rather than as the operational cleaned-filterbank baseline, its timing is reported separately from the full GPU RFI mitigation pipeline timing. For \texttt{filtool}, the reported compute time includes the mitigation computation performed by the \texttt{filtool} command, excluding file I/O under the same timing convention.

To obtain stable timing measurements, one warm-up run is performed before measurement, and each configuration is then repeated three times. We report the mean runtime over the three measured runs. For \texttt{filtool}, the reported value at each sampling time is the fastest mean runtime among the tested CPU thread counts. For the GPU RFI mitigation pipeline, all reported timings are measured using TensorRT FP16 inference, which represents the accelerated implementation used in our high-throughput pipeline. The RFDL timing is reported using the same benchmark inputs and GPU system, so that its neural-inference cost can be compared with the proposed mask predictor.

\begin{table}[t]
\centering
\caption{Compute-time benchmark protocol.}
\label{tab:runtime_protocol}
\begin{tabular}{ll}
\hline
Item & Setting \\
\hline
Compared RFI Mitigation Systems & GPU RFI mitigation pipeline, RFDL, \texttt{filtool} \\
GPU system & NVIDIA GH200 \\
CPU system & AMD EPYC 9825 \\
Input size & 4096 channels, \(\sim100\,{\rm s}\) \\
Timing type & Compute time only \\
Excluded stages & File I/O, model/engine loading, output writing, PRESTO search \\
GPU RFI mitigation pipeline timing & Full mitigation compute, including replacement and rescaling \\
RFDL timing & Neural mask-prediction compute \\
\texttt{filtool} timing & Fastest tested multi-threaded mitigation compute \\
Repeats & One warm-up run (excluded); three measured runs \\
Reported value & Mean runtime \\
\hline
\end{tabular}
\end{table}

\section{Results}
\label{sec:results}

\subsection{Patch-level RFI detection}
\label{sec:results_patch_rfi}

\begin{figure}[t]
\centering
\includegraphics[width=0.95\linewidth]{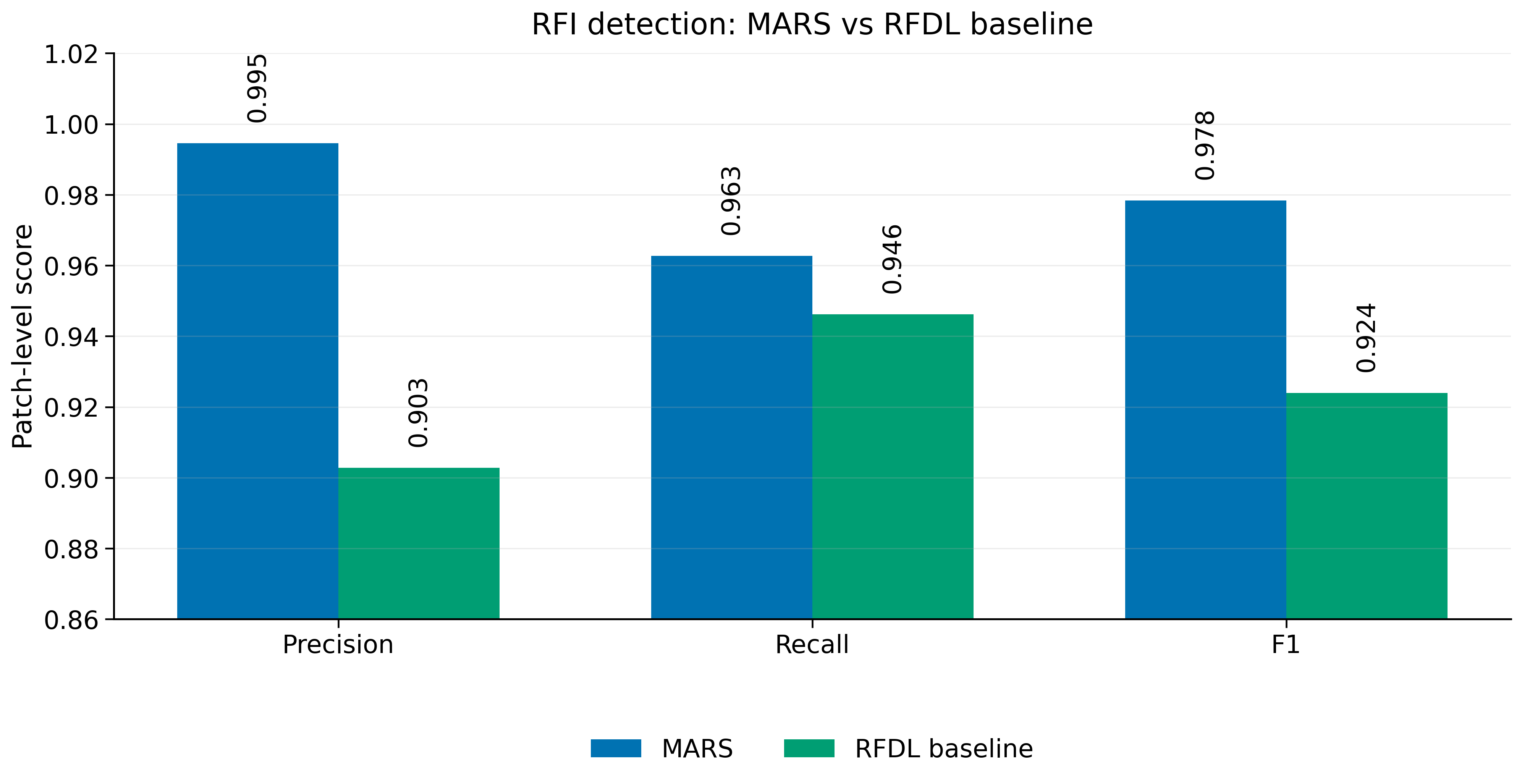}
\caption{Overall patch-level RFI detection performance on the fixed-seed synthetic RFI benchmark. MARS achieves higher precision and F1 score than the RFDL baseline when both models are evaluated using the same test patches, threshold, and metrics.}
\label{fig:rfi_detection_overall}
\end{figure}

We first evaluate patch-level RFI detection on the fixed-seed synthetic RFI benchmark described in Section~\ref{sec:synthetic_rfi_benchmark}. Figure~\ref{fig:rfi_detection_overall} compares MARS with the RFDL baseline using the same test patches and threshold. For precision, recall, and F1 score, values closer to unity indicate better performance, with the F1 score summarising the balance between precision and recall. MARS achieves a precision of \(0.995\), a recall of \(0.963\), and an F1 score of \(0.978\), compared with \(0.903\), \(0.946\), and \(0.924\), respectively, for RFDL. The proposed model therefore improves the overall F1 score while also giving a substantially higher precision, which is important because false-positive RFI masks correspond directly to clean samples being replaced in the downstream pipeline.

\begin{figure}[t]
\centering
\includegraphics[width=0.95\linewidth]{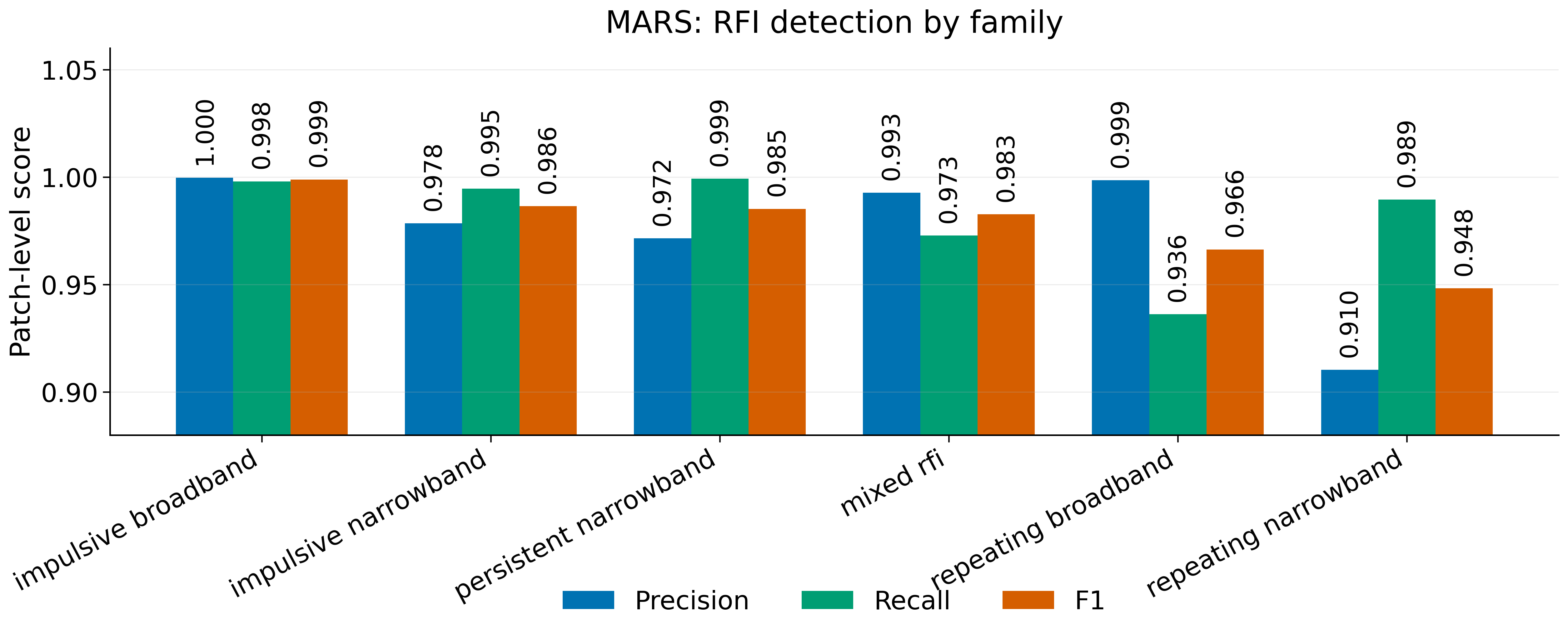}
\caption{Patch-level RFI-detection performance of MARS across the six synthetic RFI families, reported in terms of precision, recall, and F1 score.}
\label{fig:family_level MARS results}
\end{figure}

\begin{figure}[t]
\centering
\includegraphics[width=0.95\linewidth]{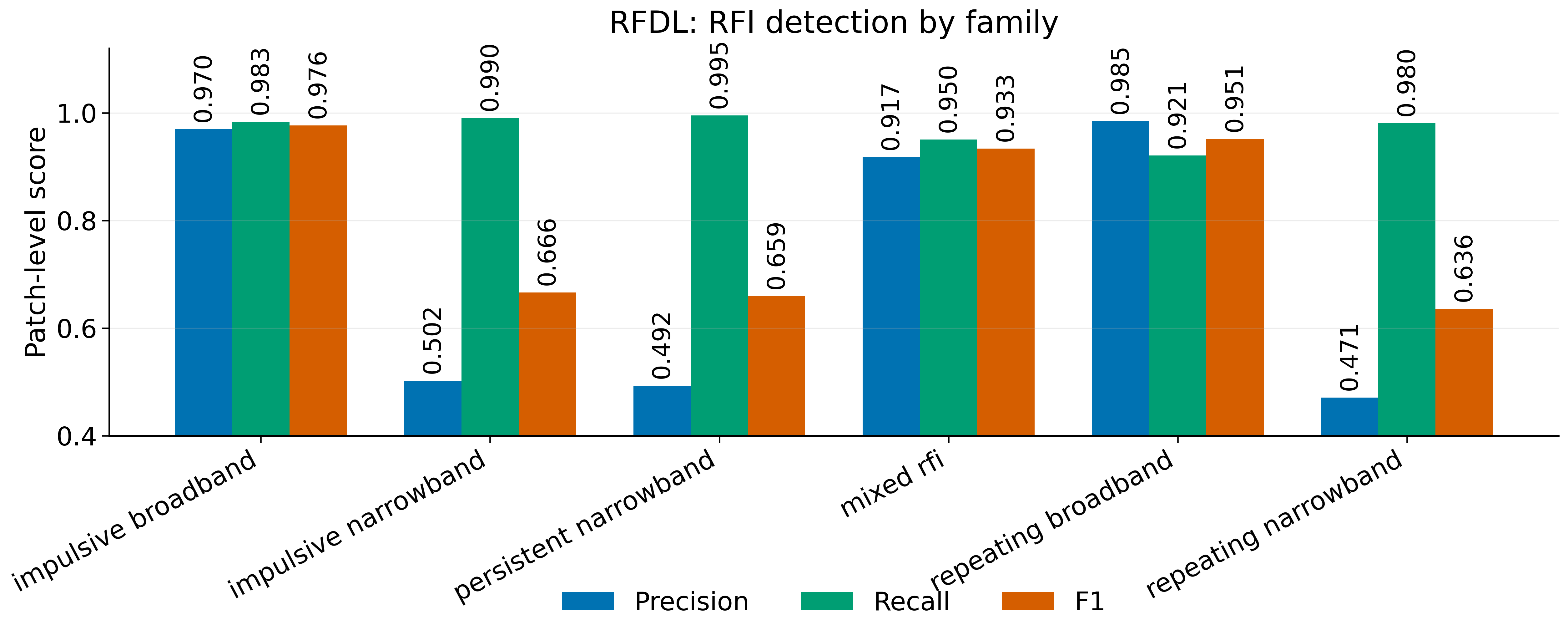}
\caption{Patch-level RFI-detection performance of RFDL across the same six synthetic RFI families, reported in terms of precision, recall, and F1 score.}
\label{fig:family_level RFDL results}
\end{figure}

Figures~\ref{fig:family_level MARS results} and~\ref{fig:family_level RFDL results} show the performance broken down by RFI family and indicate where the overall improvement arises. MARS performs well across all six families, with F1 scores of \(0.999\), \(0.986\), \(0.985\), \(0.983\), \(0.966\), and \(0.948\) for impulsive broadband, impulsive narrowband, persistent narrowband, mixed RFI, repeating broadband, and repeating narrowband RFI, respectively. The lowest F1 score occurs for repeating narrowband RFI, for which the precision decreases to \(0.910\), while the recall remains high at \(0.989\). This suggests that the model more often over-extends the predicted mask around these structures than fails to detect them.

RFDL performs competitively on broadband and mixed RFI but degrades strongly on narrowband families. It achieves F1 scores of \(0.976\), \(0.933\), and \(0.951\) for impulsive broadband, mixed RFI, and repeating broadband, respectively, but drops to \(0.666\), \(0.659\), and \(0.636\) for impulsive narrowband, persistent narrowband, and repeating narrowband RFI. In these narrowband cases, RFDL has high recall but much lower precision, indicating that it detects the contaminated structures but flags a much broader region than the ground-truth mask.

\begin{figure*}[t]
\centering
\includegraphics[width=0.49\textwidth]{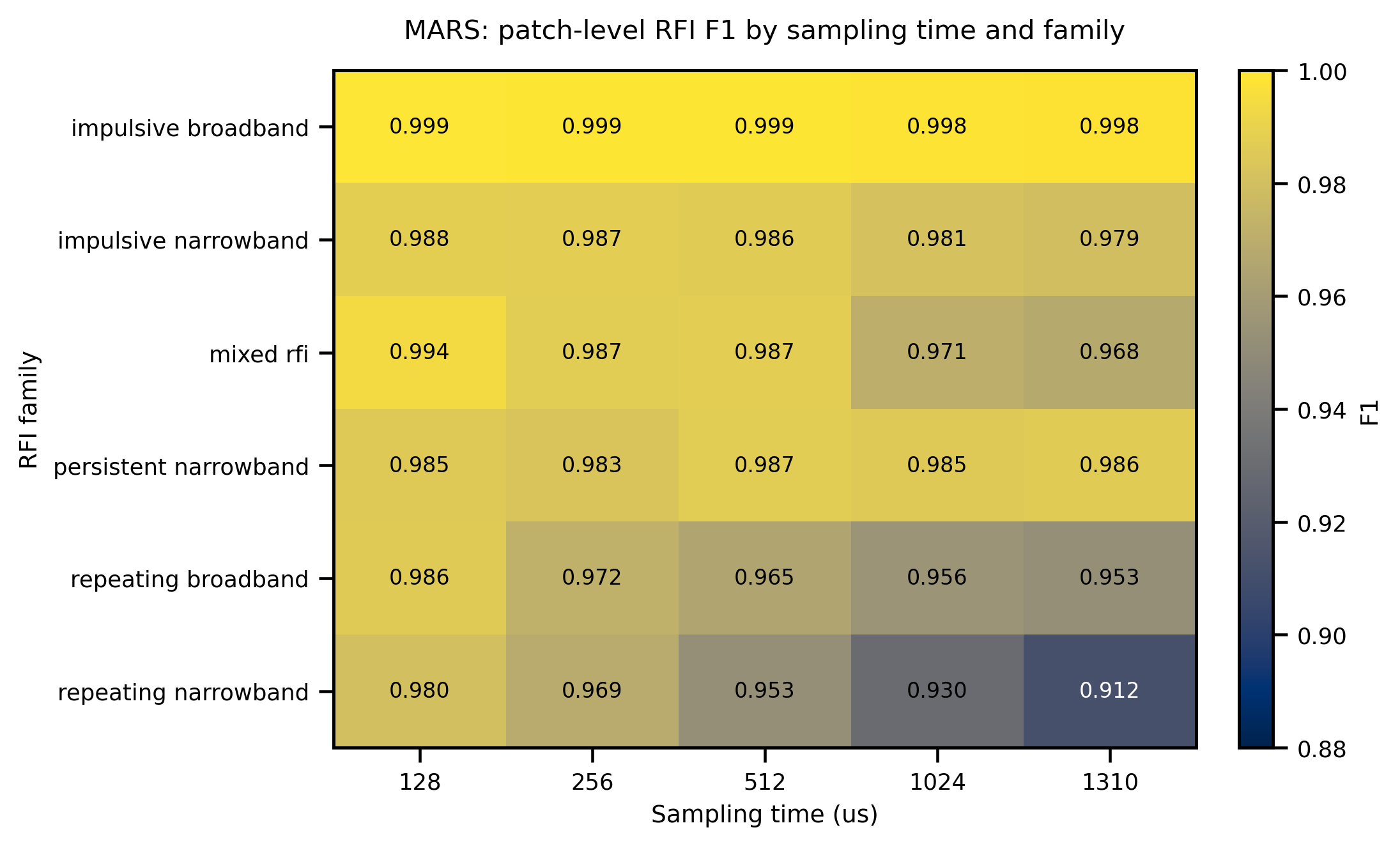}
\hfill
\includegraphics[width=0.49\textwidth]{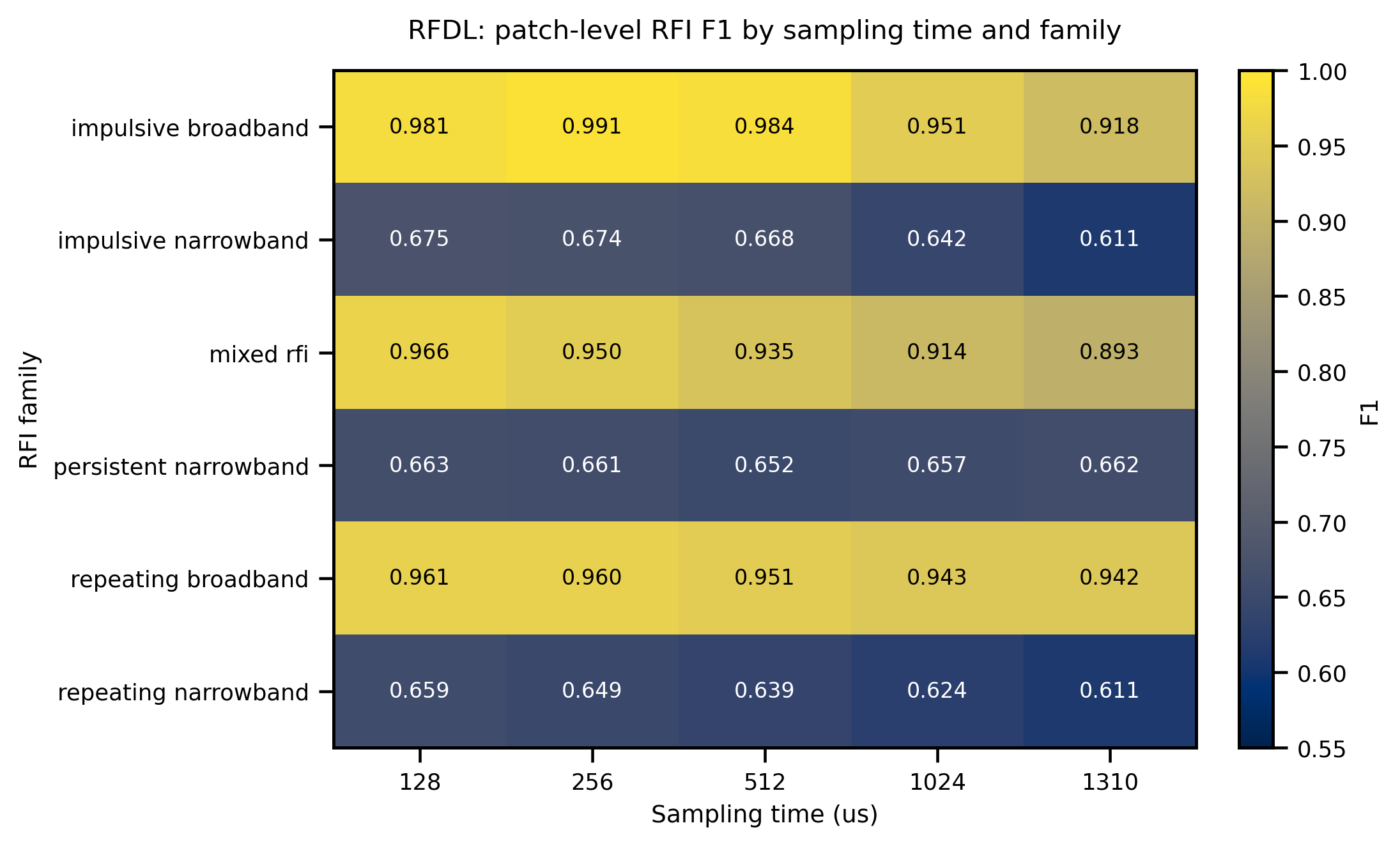}
\caption{Patch-level RFI F1 score as a function of sampling time and RFI family. Left: MARS. Right: RFDL baseline. MARS remains more stable across narrowband and mixed-RFI families, while RFDL degrades strongly for persistent and repeating narrowband structures.}
\label{fig:rfi_f1_tsamp_family}
\end{figure*}

Figure~\ref{fig:rfi_f1_tsamp_family} shows the F1 score as a function of sampling time and RFI family. RFDL remains near \(0.65\) for persistent narrowband RFI across the sampling-time grid and decreases to \(0.611\) for both impulsive and repeating narrowband RFI at \(1310\,\mu{\rm s}\). In contrast, MARS remains more stable across the grid, with an average F1 score of \(0.977\) over all family--sampling-time combinations, compared with \(0.800\) for RFDL. The clearest degradation in MARS occurs for repeating narrowband RFI at coarse sampling times, where the F1 score decreases from \(0.980\) at \(128\,\mu\mathrm{s}\) to \(0.912\) at \(1310\,\mu\mathrm{s}\). This identifies repeating narrowband RFI at coarse sampling times as the principal remaining patch-level limitation in the synthetic RFI benchmark. As indicated by the family-level results, this degradation is primarily associated with reduced precision due to mask over-extension, rather than a failure to detect the RFI structures.

Overall, the patch-level benchmark shows that MARS provides a more balanced precision-recall trade-off than RFDL and is substantially more robust to all kinds of RFI across the tested sampling-time range.

\subsection{Astrophysical-signal protection}
\label{sec:results_astro_protection}

We next evaluate astrophysical-signal protection using the fixed-seed FRB/single-pulse benchmark described in Section~\ref{sec:astro_signal_protection}. The metric is the retained fluence fraction after applying the predicted RFI mask. Larger values indicate that less injected dispersed-pulse power is removed by the RFI detector.

\begin{figure*}[t]
\centering
\includegraphics[width=0.95\textwidth]{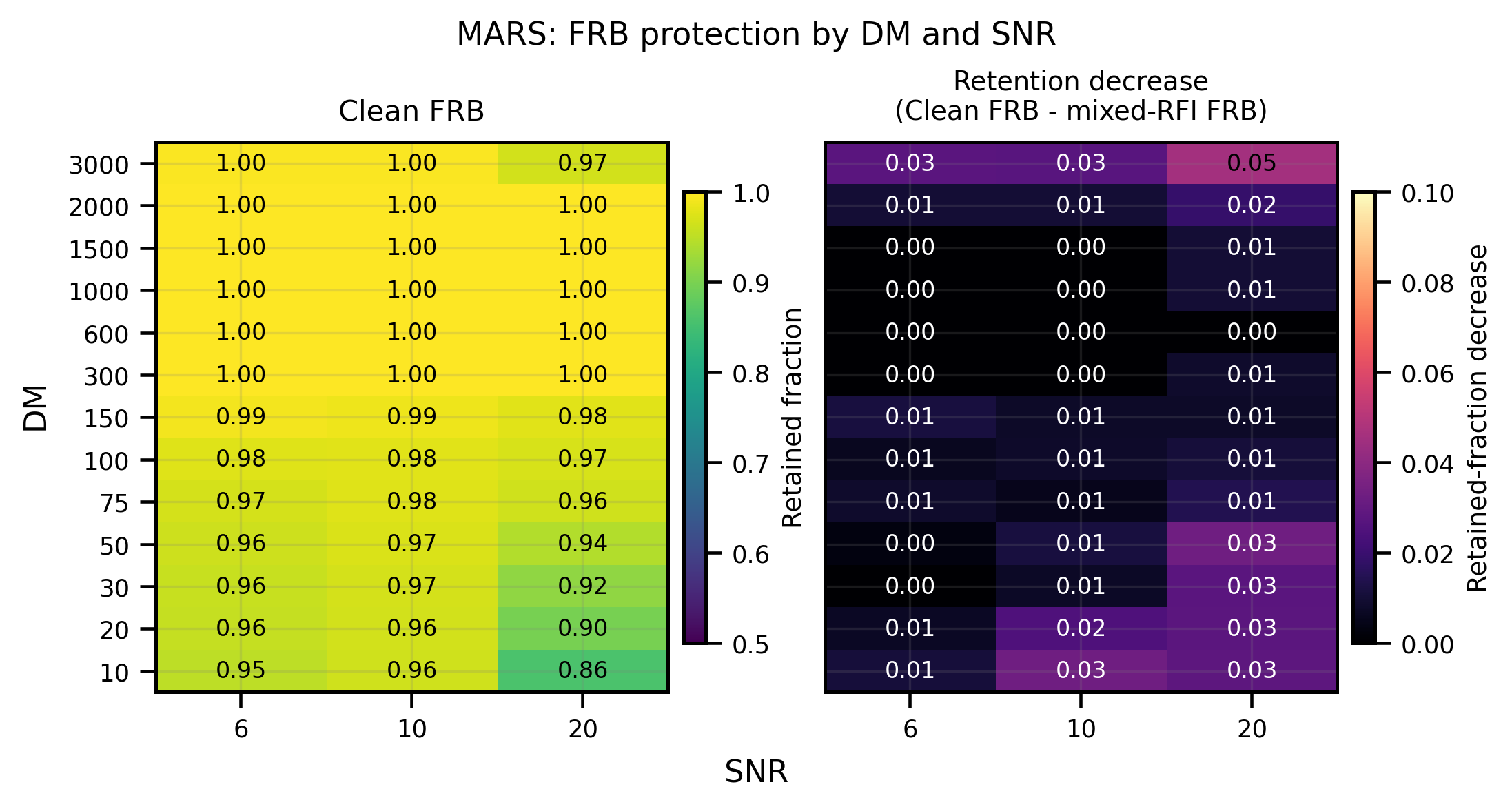}
\caption{Astrophysical-signal protection of MARS as a function of DM and S/N. The left panel shows the retained fluence fraction for clean injected FRB/single-pulse signals. The right panel shows the reduction in retained fluence caused by the addition of mixed RFI, defined as \(\Delta R=R_{\rm clean}-R_{\rm mixed\,RFI}\). Positive values therefore indicate that mixed RFI causes a larger fraction of the injected pulse to be masked, while values close to zero indicate little change in signal retention.}
\label{fig:frb_protection_grid}
\end{figure*}

Figure~\ref{fig:frb_protection_grid} shows the retained fluence of MARS as a function of DM and S/N. In the clean signal-only case, the retained fluence remains close to unity over most of the grid. The main loss occurs for low-DM, high-S/N pulses, where the dispersed track is compact and bright in the frequency--time plane and can resemble interference. Even in this difficult regime, the retained fluence remains \(0.86\) at \({\rm DM}=10\,{\rm pc\,cm^{-3}}\) and \({\rm S/N}=20\). When mixed RFI is also injected, the retained fluence remains high across most of the grid, with the lowest value decreasing to \(0.83\) at the same low-DM, high-S/N point.

\begin{figure*}[t]
\centering
\includegraphics[width=0.95\textwidth]{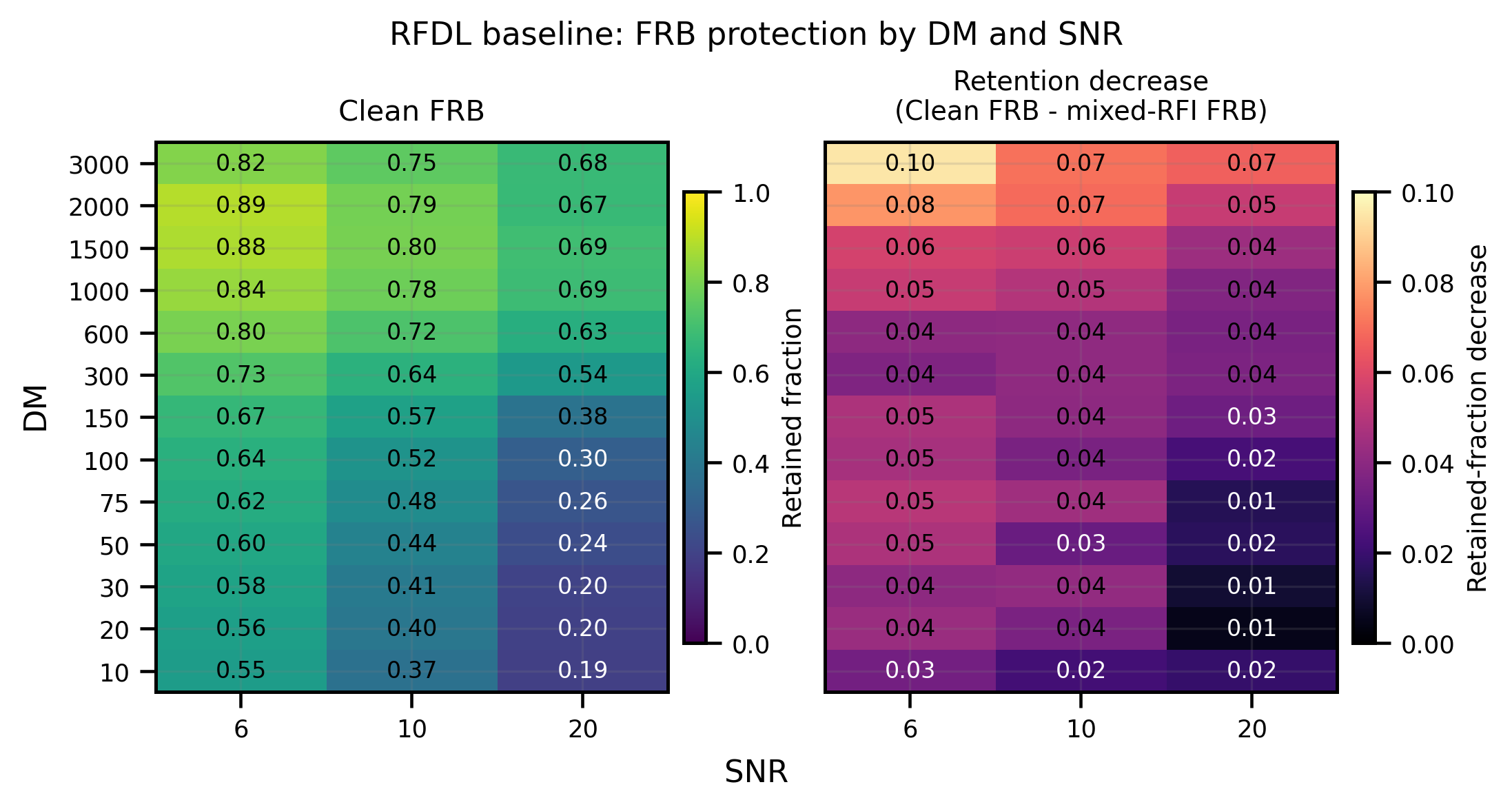}
\caption{Astrophysical-signal protection of the RFDL baseline, evaluated on the same fixed-seed FRB/single-pulse benchmark as MARS. The left panel shows the retained fluence fraction for clean injected signals. The right panel shows the reduction in retained fluence after adding mixed RFI, defined as \(\Delta R=R_{\rm clean}-R_{\rm mixed\,RFI}\). Positive values indicate additional signal loss in the presence of mixed RFI. RFDL retains less injected pulse fluence overall, particularly for low-DM, high-S/N signals.}
\label{fig:rfdl_frb_protection_grid}
\end{figure*}

\begin{figure*}[t]
\centering
\includegraphics[width=0.95\textwidth]{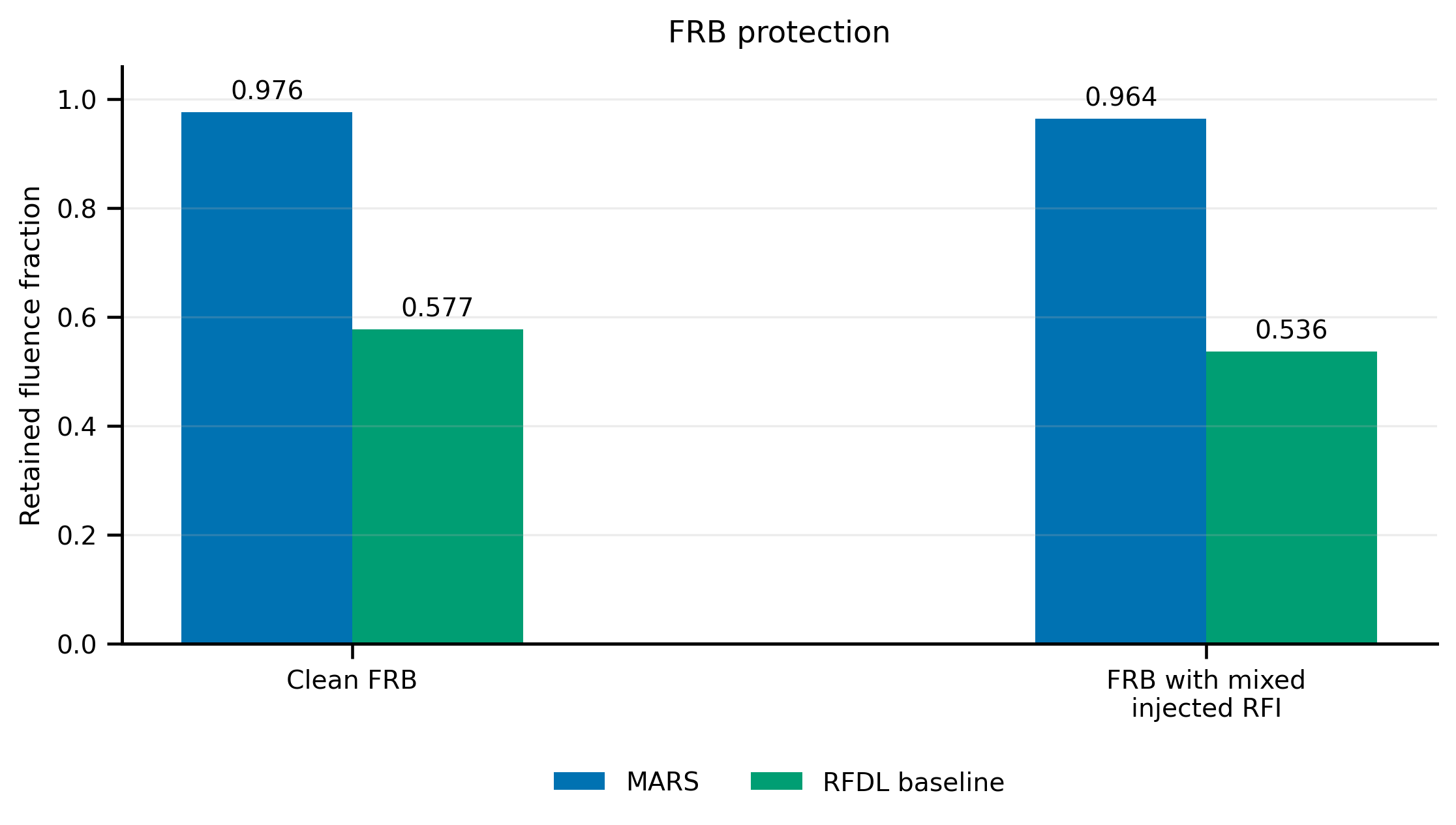}
\caption{Patch-level FRB fluence retention for MARS and RFDL, evaluated on clean FRBs and FRBs with mixed injected RFI.}
\label{fig:rfdl_frb_protection_comparison}
\end{figure*}

The comparison with RFDL shows a much larger difference in signal preservation. Figure~\ref{fig:rfdl_frb_protection_grid} shows that RFDL removes a substantial fraction of the injected dispersed-pulse fluence, especially in the low-DM and high-S/N regime. In the clean signal-only case, the retained fluence falls to \(0.19\) at \({\rm DM}=10\,{\rm pc\,cm^{-3}}\) and \({\rm S/N}=20\). With mixed injected RFI, the retained fluence at the same point decreases further to \(0.17\). As summarised in Figure~\ref{fig:rfdl_frb_protection_comparison}, the mean retained fluence over all benchmark configurations is \(0.976\) for MARS and \(0.577\) for RFDL in clean FRB patches, and \(0.964\) and \(0.536\), respectively, in patches with mixed injected RFI. This indicates that RFDL is much more likely to confuse compact, bright dispersed signals with RFI-like structures.

\begin{figure}[t]
\centering
\includegraphics[width=0.85\linewidth]{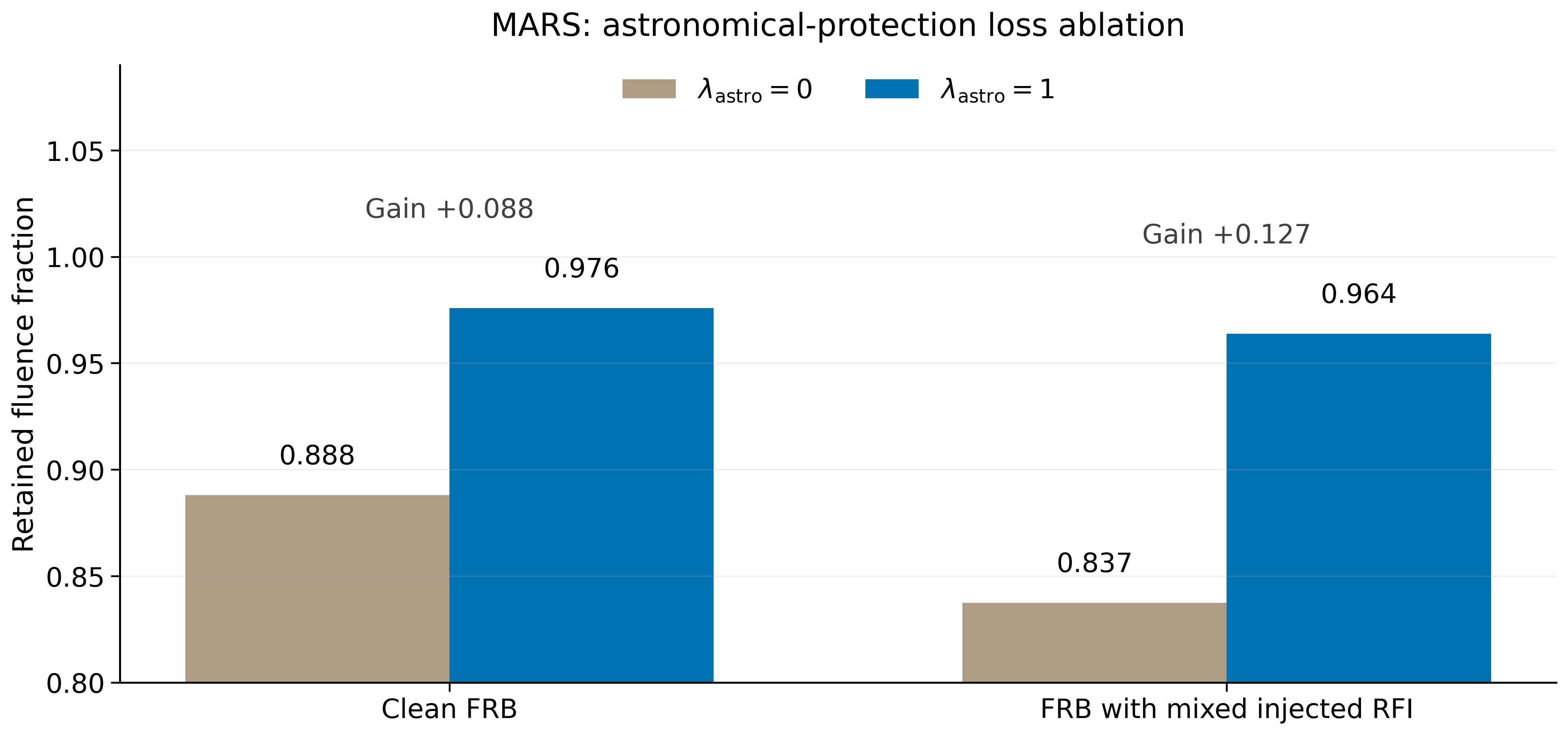}
\caption{Effect of the astronomical-signal preservation loss. The production model with \(\lambda_{\rm astro}=1\) retains substantially more injected pulse fluence than the no-\(\mathcal{L}_{\rm astro}\) ablation in both clean-FRB patches and FRB+mixed-RFI patches.}
\label{fig:lambda_astro_ablation}
\end{figure}

Figure~\ref{fig:lambda_astro_ablation} isolates the effect of the astronomical-signal preservation loss. The no-\(\mathcal{L}_{\rm astro}\) model uses the same architecture and training setup as the final model but removes the explicit preservation term from the training objective. Without this term, the retained fluence is \(0.888\) in clean signal-only patches and \(0.837\) in patches with mixed injected RFI. With \(\mathcal{L}_{\rm astro}\), the corresponding values increase to \(0.976\) and \(0.964\), giving absolute gains of \(0.088\) and \(0.127\), respectively.

\begin{figure*}[t]
\centering
\includegraphics[width=0.95\textwidth]{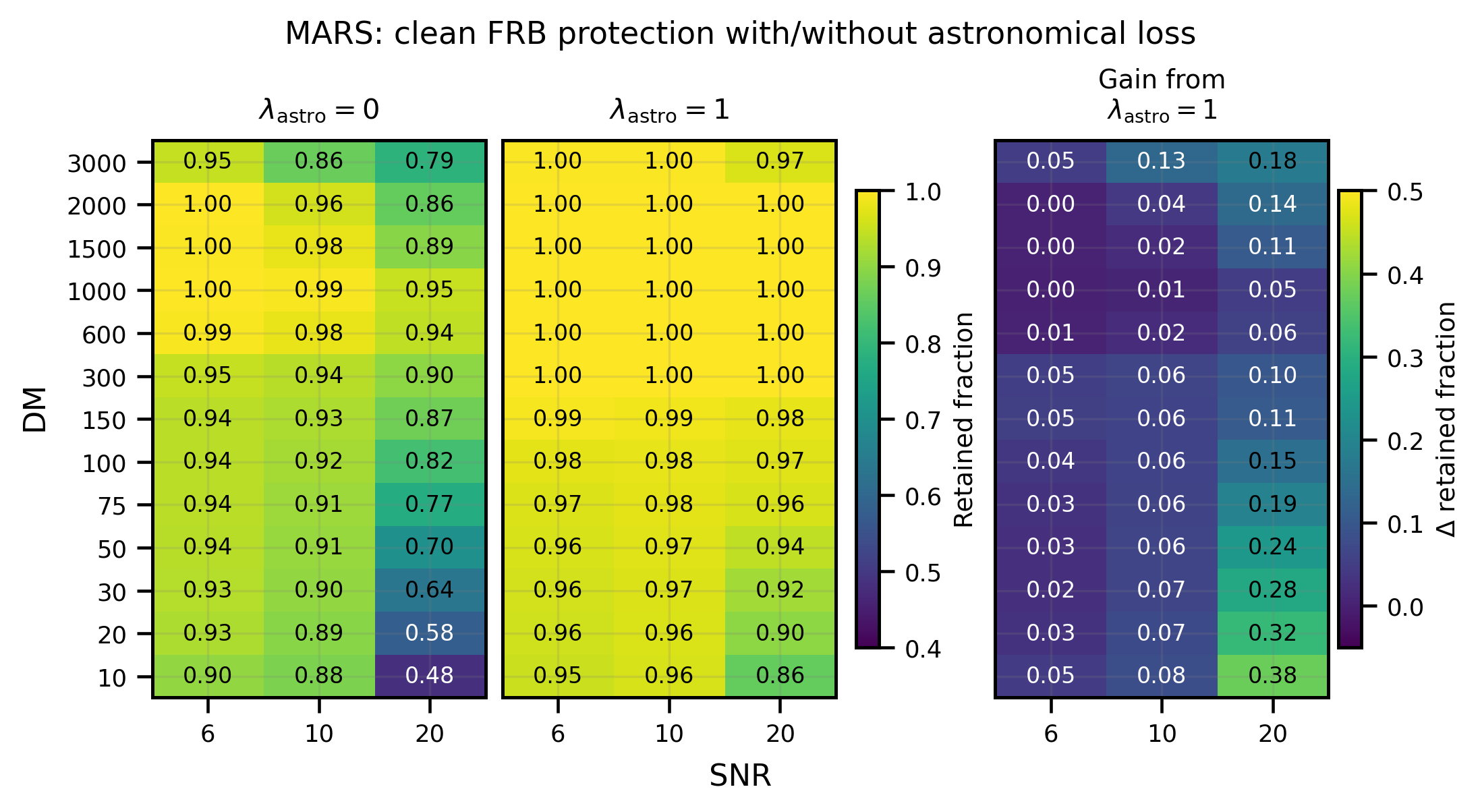}
\caption{Clean-FRB retained fluence with and without the astronomical-signal preservation loss. The largest gain from \(\lambda_{\rm astro}=1\) occurs for compact, bright dispersed signals at low DM and high S/N, where the no-\(\mathcal{L}_{\rm astro}\) model is most likely to confuse injected pulse power with RFI.}
\label{fig:lambda_astro_clean_grid}
\end{figure*}

As shown in Figure~\ref{fig:lambda_astro_clean_grid}, the improvement from \(\mathcal{L}_{\rm astro}\) is most pronounced in the low-DM, high-S/N regime. In the clean-FRB grid, the retained fluence at \({\rm DM}=10\,{\rm pc\,cm^{-3}}\) and \({\rm S/N}=20\) increases from \(0.48\) without the preservation loss to \(0.86\) with the preservation loss. This confirms that simply treating injected pulse pixels as non-RFI examples is not sufficient. The explicit preservation term is needed to discourage the network from assigning high RFI probability to compact dispersed signals that visually resemble interference.

\subsection{Synthetic full-filterbank PRESTO comparison with \texttt{filtool}}
\label{sec:results_synthetic_presto}

We next compare the GPU RFI mitigation pipeline with \texttt{filtool} on the synthetic full-filterbank PRESTO benchmark described in Section~\ref{sec:synthetic_filterbank_presto}. This is a filterbank-level comparison rather than a pixel-mask comparison. The same contaminated SIGPROC filterbank is cleaned independently by the GPU RFI mitigation pipeline and by \texttt{filtool}; both cleaned outputs are then searched with matched PRESTO settings. The metric is the recovered significance of the matched injected target candidate.

\begin{figure*}[t]
\centering
\includegraphics[width=\textwidth]{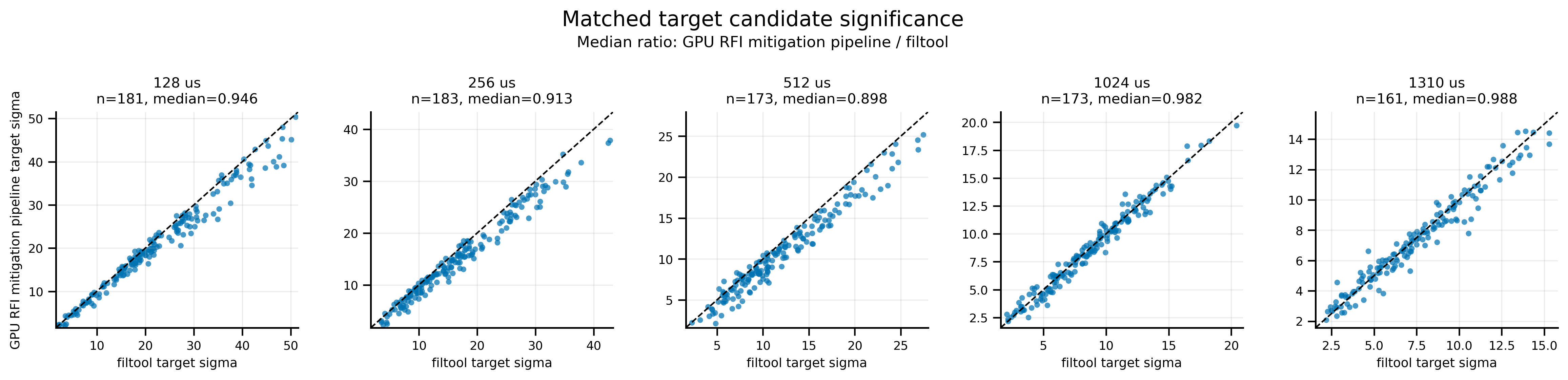}
\caption{PRESTO significance of matched target candidates recovered from synthetic contaminated filterbanks processed with \texttt{filtool} and the GPU RFI-mitigation pipeline. Each point represents one fixed-seed filterbank realisation in which the injected target signal was detected by PRESTO. In each panel, $n$ denotes the number of detected target signals included in the comparison at the corresponding sampling time. The dashed line denotes equal recovered significance; points above and below it favour the GPU pipeline and \texttt{filtool}, respectively.}
\label{fig:filtool_gpu_sigma_scatter}
\end{figure*}

\begin{figure}[t]
\centering
\includegraphics[width=0.95\linewidth]{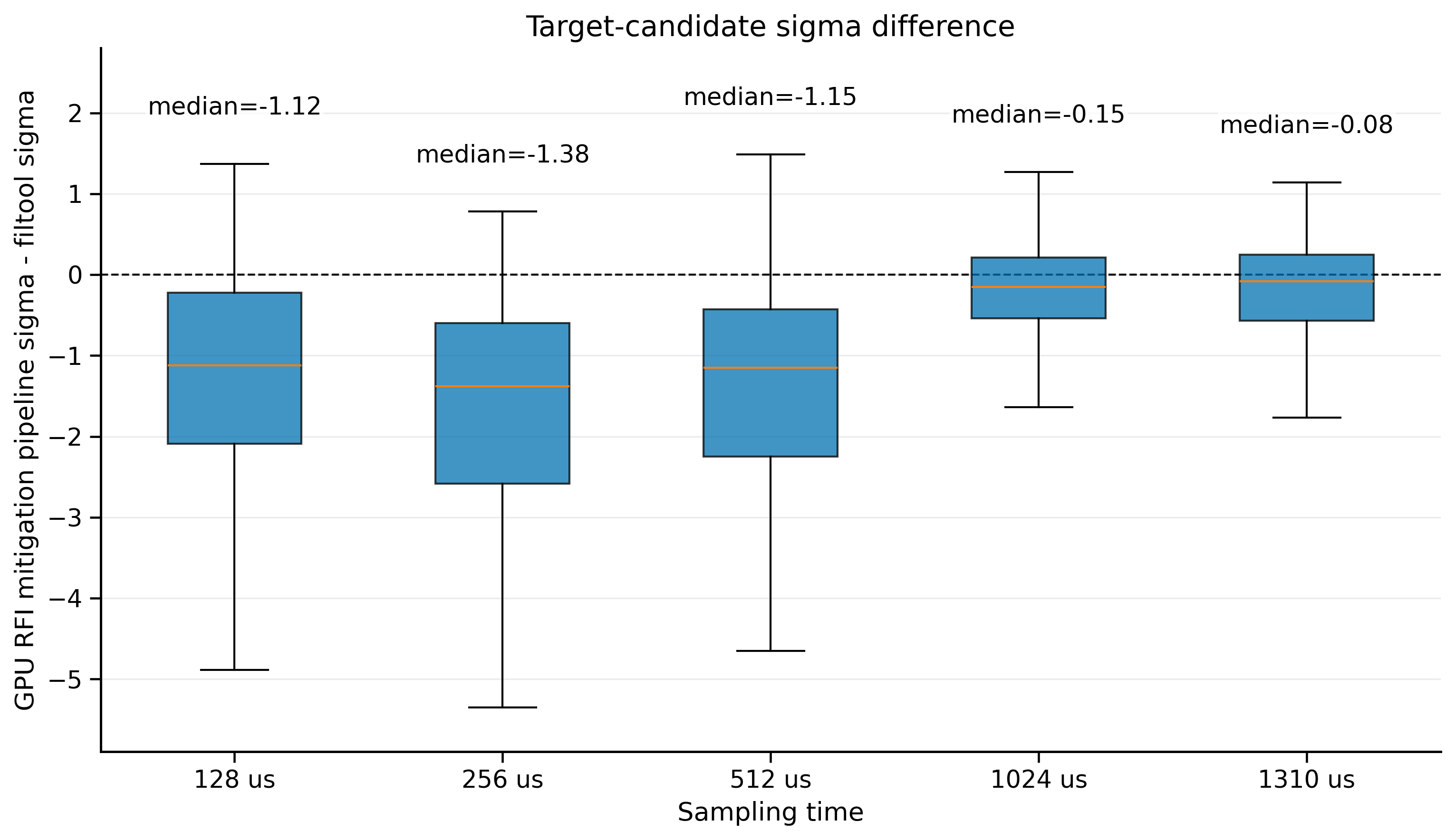}
\caption{Distribution of the difference in recovered target-candidate significance between the GPU RFI-mitigation pipeline and \texttt{filtool} at each sampling time. For candidates matched to the injected target period, we define \(\Delta\sigma=\sigma_{\rm GPU}-\sigma_{\rm filtool}\). Positive values indicate higher recovered significance with the GPU pipeline, whereas negative values indicate higher significance with \texttt{filtool}. The dashed horizontal line denotes equal significance between the two methods.}
\label{fig:filtool_gpu_sigma_delta}
\end{figure}

Figure~\ref{fig:filtool_gpu_sigma_scatter} compares the target-candidate significance recovered after GPU-NN mitigation with that recovered after \texttt{filtool} mitigation. At finer sampling times, most points lie slightly below the equality line, indicating that the GPU RFI mitigation pipeline usually recovers the target candidate with slightly lower significance than \texttt{filtool}. The median GPU-NN-to-\texttt{filtool} sigma ratios are \(0.946\), \(0.913\), and \(0.898\) at 128, 256, and \(512\,\mu{\rm s}\), respectively. The corresponding median differences in Figure~\ref{fig:filtool_gpu_sigma_delta} are \(-1.12\), \(-1.38\), and \(-1.15\).

The discrepancy becomes much smaller at coarser sampling times. At 1024 and \(1310\,\mu{\rm s}\), the median GPU-NN-to-\texttt{filtool} ratios increase to \(0.982\) and \(0.988\), and the median sigma differences are only \(-0.15\) and \(-0.08\), respectively. Thus, the GPU RFI mitigation pipeline preserves matched target-candidate significance close to that achieved by \texttt{filtool}, particularly at coarser sampling times, while operating substantially faster. This represents a small reduction in candidate significance in exchange for a substantial improvement in processing speed.

The box distributions show that the larger losses at fine sampling times are concentrated in the lower tail rather than appearing as a uniform degradation across the full test set. This is an important limitation: the GPU RFI mitigation pipeline is not identical to \texttt{filtool}, and some fine-time-resolution realisations lose more target significance after NN-based mitigation. However, the central part of the distribution remains close to the \texttt{filtool} result, and the discrepancy is small at 1024 and \(1310\,\mu{\rm s}\).

\subsection{Real GMRT observations}
\label{sec:real_gmrt_results}

We further evaluate MARS on the two real GMRT observations described in Table~\ref{tab:real_gmrt_settings}. The MARS and \texttt{filtool} outputs are searched using the same PRESTO configuration, including red-noise removal. The known pulsar signal is recovered from both mitigation outputs.

\begin{table}[t]
\centering
\caption{Recovered PRESTO significance for the known pulsar candidates in the real GMRT observations.}
\label{tab:real_gmrt_results}
\begin{tabular}{lccc}
\hline
Source & MARS & \texttt{filtool} & $\Delta\sigma$ \\
\hline
J0139+5814 & 13.89 & 13.88 & $+0.01$ \\
J0659+1414 & 17.09 & 16.90 & $+0.19$ \\
\hline
\end{tabular}
\end{table}

As shown in Table~\ref{tab:real_gmrt_results}, MARS preserves the known periodic signals at significance values comparable to, and slightly higher than, those obtained with \texttt{filtool}. These observations provide an initial real-data validation, although a larger sample is required for a broader comparison.

\subsection{RFI-mitigation compute-time comparison}
\label{sec:results_compute_time}

We compare the compute-time performance of the GPU RFI mitigation pipeline, the RFDL neural baseline, and \texttt{filtool} using the protocol in Section~\ref{sec:compute_time_benchmark}. The GPU RFI mitigation pipeline and \texttt{filtool} are operational mitigation pipelines, whereas RFDL is timed as a neural mask-prediction baseline. All reported times exclude file I/O, model or engine loading, output writing, and downstream PRESTO searching.

\begin{figure*}[t]
\centering
\includegraphics[width=\textwidth]{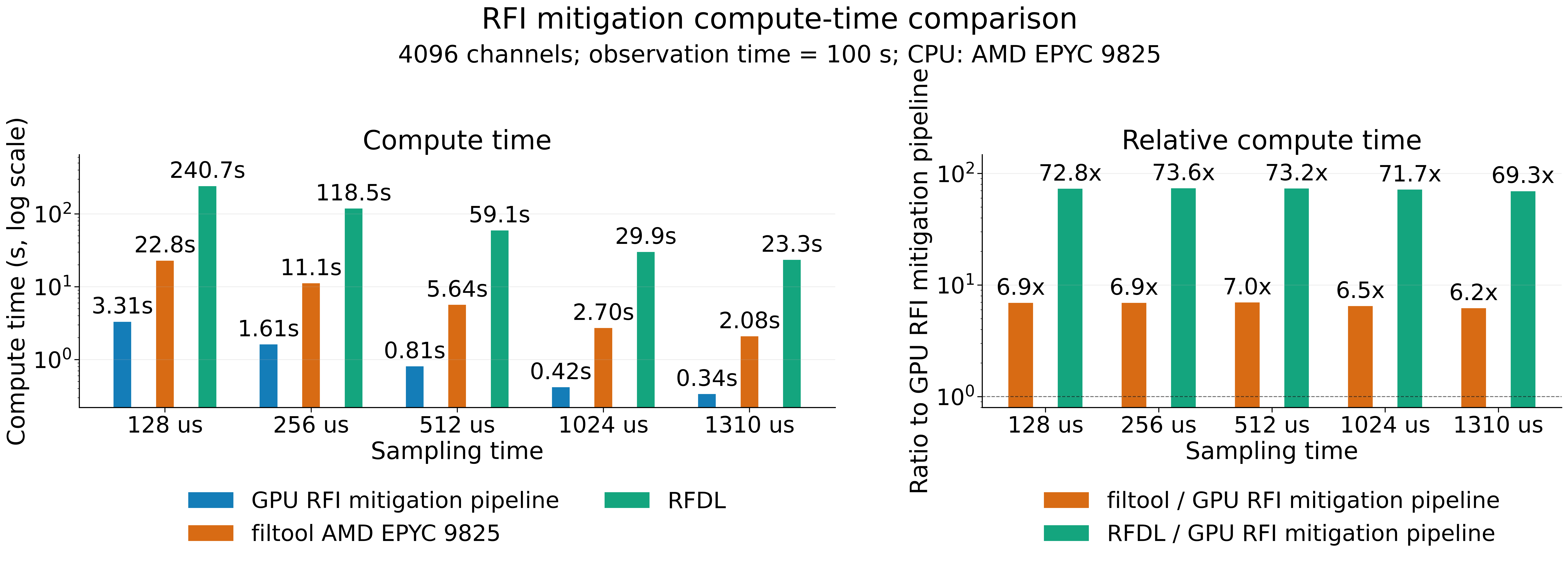}
\caption{Compute-only runtime comparison among the GPU RFI mitigation pipeline, RFDL, and \texttt{filtool}. Left: absolute compute time for 4096-channel, \(\sim100\,{\rm s}\) filterbanks. Right: compute time normalised by the GPU RFI mitigation pipeline. The GPU RFI mitigation pipeline and \texttt{filtool} timings correspond to operational mitigation compute time, while RFDL is included as a neural mask-prediction speed baseline. File I/O, model or engine loading, output writing, and downstream PRESTO processing are excluded.}
\label{fig:compute_time_comparison}
\end{figure*}

Figure~\ref{fig:compute_time_comparison} shows that the GPU RFI mitigation pipeline is the fastest method at every tested sampling time. For 4096-channel, \(\sim100\,{\rm s}\) filterbanks, the GPU RFI mitigation pipeline completes the full mitigation computation in \(0.34\)--\(3.31\,{\rm s}\), while \texttt{filtool} requires \(2.08\)--\(22.8\,{\rm s}\). This corresponds to a \(6.2\)--\(7.0\times\) reduction in mitigation compute time relative to the fastest tested multi-threaded \texttt{filtool} configuration.

The comparison with RFDL is even more restrictive in favour of the baseline, because the RFDL timing includes only the computation needed to produce the inference mask, whereas the GPU RFI mitigation pipeline timing includes preprocessing, neural inference, and post-processing. Even under this comparison, RFDL requires \(23.3\)--\(240.7\,{\rm s}\), or \(69.3\)--\(73.6\times\) the GPU RFI mitigation pipeline runtime. Thus, the proposed pipeline completes the entire RFI-removal computation much faster than RFDL produces a mask alone.

These results demonstrate that the speed advantage stems not merely from executing the neural network model on a GPU, but also from the lightweight MARS design, TensorRT optimisation, and a GPU implementation of the complete filterbank processing pipeline. Under a timing protocol that accounts only for computational time, the execution time of this GPU-based neural network pipeline is significantly lower than that of traditional methods. Consequently, this approach enables real-time or near-real-time RFI mitigation while maintaining the detection and signal preservation performance described earlier.

\section{Discussion}
\label{sec:discussion}

\subsection{Balancing RFI removal and astronomical-signal preservation}
\label{sec:discussion_signal_preservation}

The main result of this work is that RFI mitigation cannot be judged by RFI-mask recall alone. A useful mitigation method must also avoid replacing clean astrophysical signal power. In the synthetic patch benchmark, MARS improves the overall F1 score from \(0.924\) for RFDL to \(0.978\), with a particularly large gain in precision. This precision gain is important because false-positive RFI masks become replaced samples in the downstream filterbank pipeline.

The FRB-protection benchmark shows the same point more directly. MARS retains \(0.976\) of the injected fluence in clean signal-only patches and \(0.964\) in FRB+mixed-RFI patches, whereas RFDL retains only \(0.577\) and \(0.536\), respectively. The gap is largest for low-DM, high-S/N pulses, where compact dispersed tracks resemble interference most closely. This indicates that a model can perform reasonably as an RFI detector while still being unsuitable for science-preserving mitigation if it masks bright dispersed pulse power.

The astronomical-signal preservation loss is a key contributor to this behaviour. Removing \(\mathcal{L}_{\rm astro}\) reduces retained fluence from \(0.976\) to \(0.888\) in clean patches and from \(0.964\) to \(0.837\) in FRB+mixed-RFI patches. The largest gain occurs in the low-DM, high-S/N regime. Thus, injecting pulse-like examples during training is useful, but the explicit preservation term is needed to penalise high RFI probability on clean dispersed-signal pixels.

At the filterbank level, the PRESTO comparisons show that the GPU RFI mitigation pipeline generally preserves matched target-candidate significance close to the \texttt{filtool} baseline. The synthetic full-filterbank tests show a modest lower-tail degradation at finer sampling times, but the median discrepancy becomes small at 1024 and \(1310\,\mu{\rm s}\). The two real GMRT checks also recover the known pulsars with comparable or slightly higher target significance after GPU-NN mitigation. These results support the interpretation that the proposed method balances RFI removal with signal preservation in the tested settings.

\subsection{Comparison with traditional and neural baselines}
\label{sec:discussion_baselines}

The two baselines test different aspects of the mitigation problem. RFDL is a neural mask-prediction baseline and is therefore compared at patch level. \texttt{filtool} is an operational filterbank-mitigation tool and is therefore compared at the cleaned-filterbank level through matched PRESTO searches. This separation is important: RFDL tests whether the proposed neural detector is a better mask predictor, while \texttt{filtool} tests whether the complete GPU RFI mitigation pipeline produces search-compatible cleaned filterbanks.

Against RFDL, the proposed model gives both better RFI-mask performance and much stronger astrophysical-signal protection. The narrowband RFI families are the clearest mask-level difference: RFDL often detects these structures with high recall but low precision, whereas MARS maintains a more balanced precision--recall trade-off. The FRB-protection results show an even larger difference, indicating that RFDL is more likely to treat compact dispersed signals as interference.

Against \texttt{filtool}, the comparison is more operational. The GPU RFI mitigation pipeline does not reproduce the \texttt{filtool} output exactly, and at fine sampling times some synthetic realisations show lower recovered target significance. However, the median PRESTO recovery remains close to \texttt{filtool}, especially at coarser sampling times, while the GPU RFI mitigation pipeline substantially reduces mitigation compute time. This suggests that the proposed method is best viewed as a fast GPU-compatible mitigation alternative rather than a replica of \texttt{filtool}.

The runtime comparison further shows that the speed advantage is not merely due to using a neural model on GPU. RFDL is also a neural baseline, but the time required for RFDL to produce an inference mask alone is \(69.3\)--\(73.6\times\) longer than the time required for the GPU RFI mitigation pipeline to complete the full mitigation computation. The proposed pipeline is also \(6.2\)--\(7.0\times\) faster than the fastest tested multi-threaded \texttt{filtool} configuration under the compute-only timing protocol. This supports the design choice of using a small TensorRT-oriented mask predictor embedded in a GPU processing pipeline.

\subsection{Limitations and future work}
\label{sec:discussion_limitations}

Several limitations remain. First, the real-data validation is currently limited to two GMRT observations. These tests show that the GPU RFI mitigation pipeline does not suppress the known target pulsars in the tested cases, but they do not establish robustness across telescopes, receivers, observing bands, or RFI environments. A larger real-observation sample is required before making broader claims about real-data generalisation.

Second, most quantitative results are based on controlled injection tests. The SPECTRALib patch benchmark, FRB-protection benchmark, and synthetic full-filterbank PRESTO tests provide known ground truth and reproducible comparisons, but they cannot fully capture the diversity of real instrumental effects and RFI morphologies. This simulation-to-real gap remains an important source of uncertainty.

Third, the synthetic filterbank PRESTO comparison shows a lower-tail degradation at finer sampling times. At 128, 256, and \(512\,\mu{\rm s}\), most GPU-NN-cleaned cases remain close to the \texttt{filtool} result, but a subset loses more target significance. This may reflect mask-boundary errors or replacement behaviour that affects the downstream Fourier search.

Fourth, low-DM, high-S/N dispersed signals remain the most difficult astrophysical signals to protect. The astronomical-signal preservation loss substantially improves retained fluence in this regime, but compact bright dispersed tracks are still more easily confused with RFI-like structures than higher-DM or lower-S/N signals.

Finally, future work should include larger real-data validation and memory-aware chunking for longer filterbanks that cannot be processed in a single GPU-resident pass.

\section{Summary and Conclusions}
\label{sec:conclusion}

We have presented MARS, a lightweight morphology-aware neural network for GPU-based RFI mitigation. The model identifies RFI in frequency-time data, while the complete GPU pipeline performs mask-guided replacement and output conditioning. The training objective includes an astronomical-signal preservation term designed to reduce the false masking of dispersed astrophysical pulses.

The proposed model outperforms the RFDL neural-network baseline on the SPECTRALib patch-level RFI benchmark, improving the overall F1 score from \(0.924\) to \(0.978\) and the precision from \(0.903\) to \(0.995\). It also provides substantially better protection of injected dispersed signals, retaining \(0.976\) of the signal fluence in clean patches and \(0.964\) in signal-plus-RFI patches. The ablation study confirms that \(\mathcal{L}_{\rm astro}\) is an important component of this signal-preservation performance.

At filterbank level, the proposed NN pipeline produces downstream PRESTO-search target-candidate significance close to the traditional \texttt{filtool} baseline. In two real GMRT observations, J0659+1414 and J0139+5814, the known pulsars are recovered with comparable or slightly higher significance after NN-based mitigation. The runtime benchmark further shows that the NN pipeline is substantially faster than the fastest tested multi-threaded \texttt{filtool} configuration, with compute-only speedups between \(6.2\times\) and \(7.0\times\).

Overall, these results show that the proposed pipeline provides a practical GPU-accelerated alternative for RFI mitigation in pulsar-search workflows, combining accurate RFI masking, explicit astrophysical-signal protection, downstream PRESTO-search compatibility, and reduced mitigation runtime. Future work will extend the validation to larger real-observation samples and develop memory-aware processing strategies for longer filterbanks.
%
\facilities{GMRT, Isambard-AI}

\software{
SIGPROC \citep{lorimer2011sigproc},
sigpyproc \citep{sigpyproc_software},
SPECTRALib \citep{spectralib_software},
PyTorch \citep{paszke2019pytorch},
TensorFlow \citep{tensorflow2015-whitepaper},
PulsarX/\texttt{filtool} \citep{men2023pulsarx},
PRESTO \citep{ransom2002fourier},
TensorRT \citep{nvidia_tensorrt_docs}
}


\clearpage
\appendix

\section{Decoder-refinement ablation study}
\label{app:decoder_ablation}

We perform an ablation study to evaluate the contribution of the decoder-refinement components in MARS. For each ablation variant, the corresponding refinement component is removed before training, and the modified model is retrained from scratch using the same training protocol as MARS. The experiment therefore measures the performance of independently trained architectures rather than the effect of disabling a component only during inference.

Four configurations are compared: MARS, a model without the vertical refinement branch, a model without the horizontal refinement branch, and a model without either decoder-refinement branch. Within each benchmark, all variants are evaluated using the same fixed test set and inference settings. We report the RFI F1 score obtained from pixel counts pooled over the 15,000 test patches, the retained FRB fluence in 9360 patches containing mixed injected RFI, and the recall measured over RFI pixels that do not overlap the injected FRB. The retained-fluence metric is similarly evaluated only over injected-pulse support that is not physically overlapped by RFI. Hysteresis post-processing is disabled for all variants. The aggregate ablation results are summarised in Table~\ref{tab:decoder_ablation_summary}.

\begin{table}[H]
\centering
\caption{Summary of the decoder-refinement ablation study. Each ablated model is retrained after removing the corresponding refinement component. RFI F1 is calculated from aggregate pixel counts, and FRB retention is measured over pulse support not physically overlapped by injected RFI.}
\label{tab:decoder_ablation_summary}
\begin{tabular}{lccc}
\hline
Model variant & RFI F1 & FRB fluence retained & RFI-only recall \\
\hline
MARS                 & 0.978 & 0.964 & 0.992 \\
No vertical refine   & 0.969 & 0.935 & 0.989 \\
No horizontal refine & 0.985 & 0.928 & 0.990 \\
No decoder refine    & 0.969 & 0.888 & 0.986 \\
\hline
\end{tabular}
\end{table}

\FloatBarrier

\begin{figure}[H]
\centering
\includegraphics[width=0.98\linewidth]{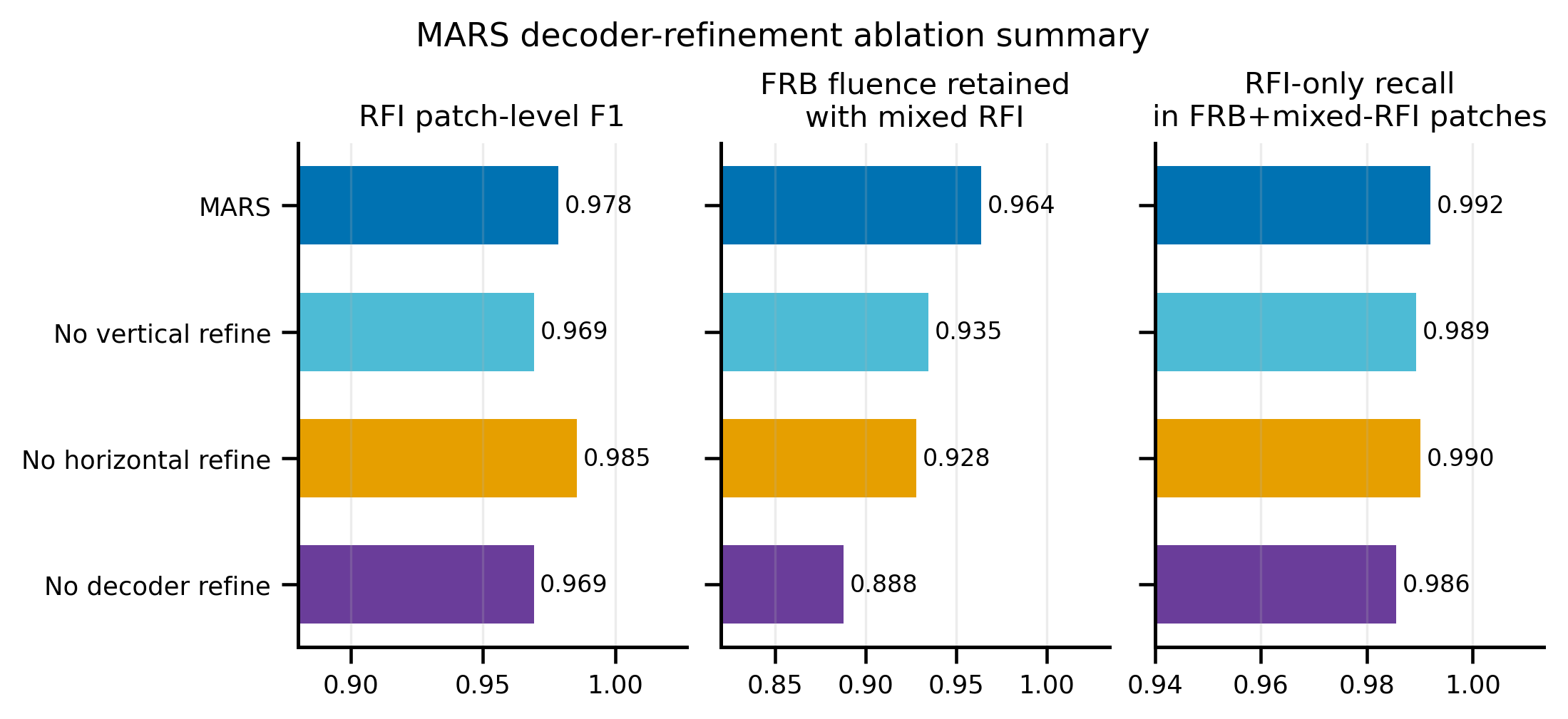}
\caption{Summary of the decoder-refinement ablation study. The panels compare aggregate patch-level RFI F1, retained FRB fluence in patches containing mixed injected RFI, and RFI-only recall in FRB-plus-mixed-RFI patches. FRB and RFI metrics exclude pixels where the two injected signals physically overlap.}
\label{fig:decoder_ablation_summary}
\end{figure}

\FloatBarrier

Figure~\ref{fig:decoder_ablation_summary} shows that the refinement branches introduce a trade-off between RFI segmentation and astrophysical-signal preservation. MARS achieves an aggregate RFI F1 score of \(0.978\), a retained FRB fluence fraction of \(0.964\), and an RFI-only recall of \(0.992\). Removing the vertical refinement branch changes these values to \(0.969\), \(0.935\), and \(0.989\), respectively. Removing the horizontal refinement branch increases the aggregate RFI F1 score to \(0.985\), but reduces retained FRB fluence to \(0.928\), with an RFI-only recall of \(0.990\). Removing both decoder-refinement branches gives an RFI F1 score of \(0.969\), while producing the lowest retained FRB fluence, \(0.888\), and an RFI-only recall of \(0.986\). The complete refinement design is therefore retained because it gives the strongest FRB preservation while maintaining high RFI-segmentation performance.

\begin{figure}[H]
\centering
\includegraphics[width=0.98\linewidth]{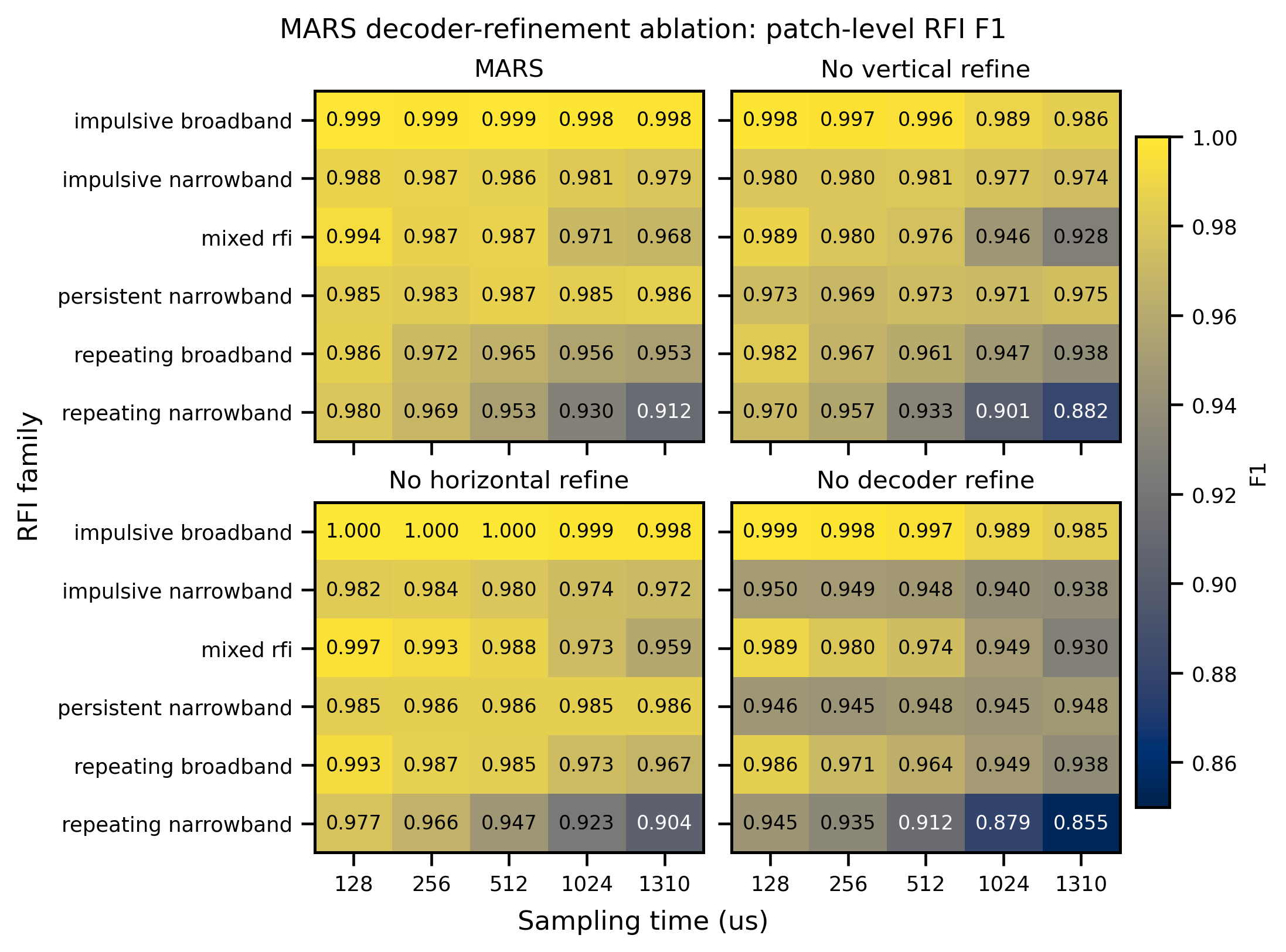}
\caption{Aggregate patch-level RFI F1 scores of the decoder-refinement ablation variants as functions of sampling time and RFI family. Each model is evaluated on the same family-stratified test set.}
\label{fig:decoder_ablation_rfi_heatmap}
\end{figure}

\FloatBarrier

The stratified results in Figure~\ref{fig:decoder_ablation_rfi_heatmap} show that the effect of decoder refinement depends on RFI morphology and sampling time. Repeating narrowband RFI at coarse sampling times is challenging for all variants. At \(1310\,\mu{\rm s}\), the repeating-narrowband F1 score is \(0.912\) for MARS, compared with \(0.882\) without vertical refinement, \(0.904\) without horizontal refinement, and \(0.855\) without decoder refinement. Although removing the horizontal branch improves the aggregate F1 score, MARS is more reliable for this difficult family and provides substantially better FRB preservation.

\begin{figure}[H]
\centering
\includegraphics[width=0.98\linewidth]{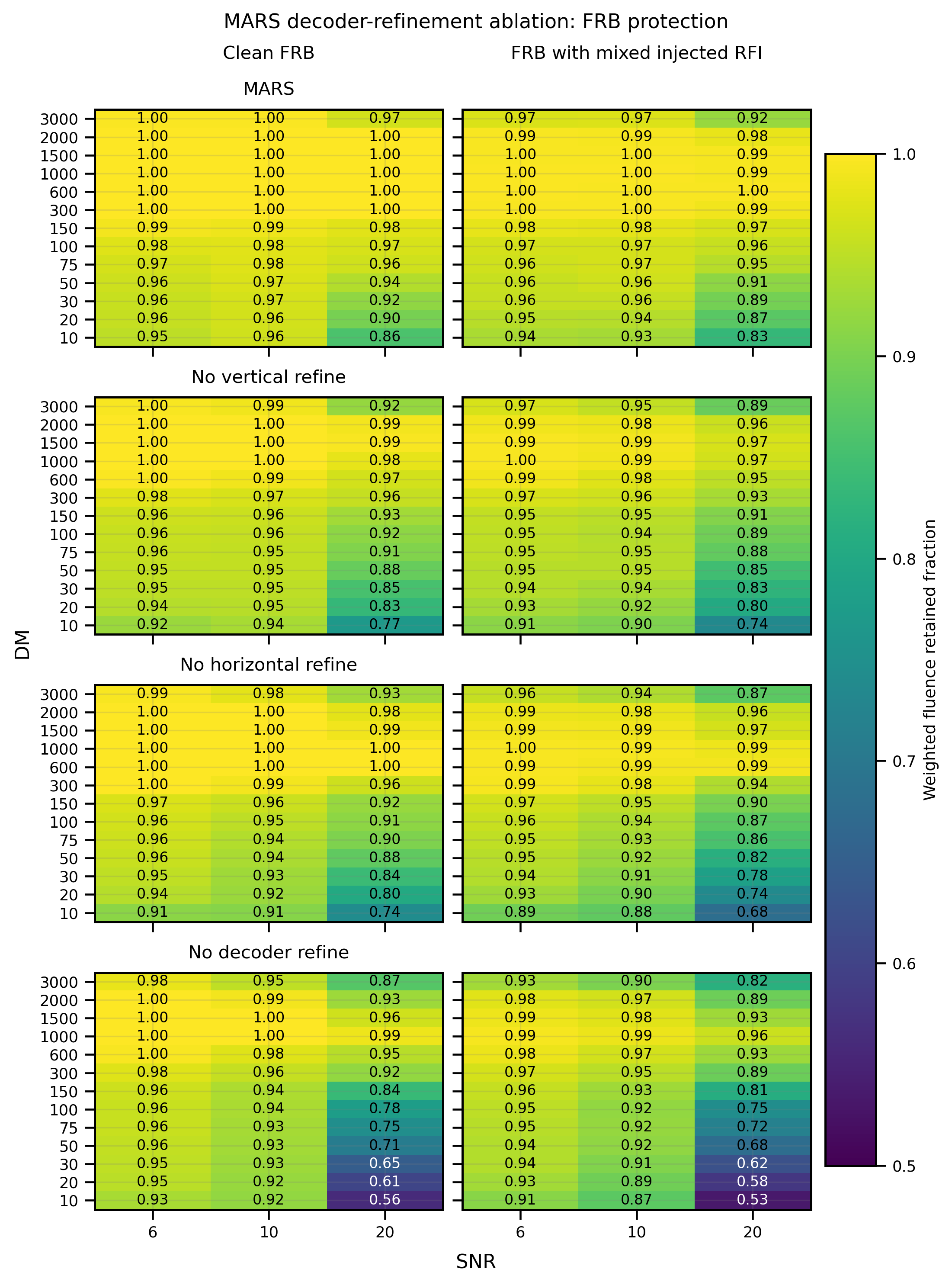}
\caption{Astrophysical-signal protection of the decoder-refinement ablation variants. For each variant, the left panel shows retained fluence for clean injected FRB signals, while the right panel shows retained fluence when mixed SPECTRALib RFI is also injected. In the mixed-RFI benchmark, retention is evaluated over pulse support that is not physically overlapped by RFI.}
\label{fig:decoder_ablation_frb_heatmap}
\end{figure}

\FloatBarrier

Figure~\ref{fig:decoder_ablation_frb_heatmap} shows that the decoder-refinement components also affect astrophysical-signal preservation. The largest differences occur in the low-DM, high-S/N regime, where compact bright dispersed signals are most easily confused with RFI. At DM \(=10\,{\rm pc\,cm^{-3}}\) and S/N \(=20\) in the FRB-plus-mixed-RFI benchmark, MARS retains \(0.83\) of the injected fluence, compared with \(0.74\) without vertical refinement, \(0.68\) without horizontal refinement, and \(0.53\) without decoder refinement. The no-decoder-refine variant is therefore the weakest in this regime. Overall, the ablation indicates that the complete refinement design provides the most favourable balance between RFI segmentation and preservation of dispersed astrophysical signals.

\section{RFDL baseline training configuration}
\label{app:rfdl_training_config}

The RFDL baseline is retrained for this work rather than evaluated using an off-the-shelf checkpoint. To provide a controlled neural-network baseline comparison, RFDL and MARS use the same \(512\times512\) source patches and the same training and validation split:
\[
N_{\rm train}=4800,\qquad N_{\rm val}=1200.
\]
MARS operates directly on the \(512\times512\) patches, whereas the native RFDL pipeline divides each source patch into \(64\times64\) sub-patches after preprocessing. Although the two implementations are maintained in separate repositories, the duplicated training and validation arrays were verified to have matching shapes, file sizes, and sampled content hashes. They are therefore treated as the same source-data split.

The RFI and FRB augmentation distributions are also matched between the two models. Both use the same random seed, time-resolution set, RFI-family sampling probabilities, and injected FRB parameter ranges. The controlled settings are summarised in Table~\ref{tab:rfdl_fair_training_config}.

\begin{table}[H]
\centering
\caption{Controlled source-data and augmentation settings used for both MARS and the RFDL baseline.}
\label{tab:rfdl_fair_training_config}
\begin{tabular}{ll}
\hline
Component & Setting \\
\hline
Training source patches & 4800 \\
Validation source patches & 1200 \\
Source patch size & \(512\times512\) \\
Random seed & 1234 \\
Training augmentation & Enabled \\
Validation augmentation & Disabled \\
Sampling times & 128, 256, 512, 1024, \(1310\,\mu{\rm s}\) \\
Maximum RFI families per patch & 3 \\
Extra-family probability & 0.35 \\
FRB pulse-width range & 1--10 ms \\
Low-DM FRB pulse-width range & 1--5 ms \\
Low-DM bright-medium pulse-width range & 3--6 ms \\
RFI-family definitions & Matched \\
RFI-family sampling weights & Matched \\
Positive-mask weights & Matched \\
\hline
\end{tabular}
\end{table}

The principal difference between the two training pipelines is the native input representation required by each model. MARS is trained directly on the normalised, tanh-compressed \(512\times512\) patches used by the GPU mitigation pipeline. RFDL instead expects positive raw-like amplitudes before applying its own logarithmic and min--max preprocessing. For RFDL training, the shared morphology augmentation is first applied to the tanh-domain source patch. The augmented patch is then mapped to the positive raw-like amplitude range required by RFDL, processed using the native RFDL transformation, and divided into \(64\times64\) sub-patches.

RFDL is trained in two stages. First, its convolutional denoising autoencoder (CDAE) is trained to reconstruct a clean-like representation of each contaminated sub-patch. The smoothed CDAE output is then subtracted from the contaminated input, and the resulting residual is used to train the RFDL mask detector. The model-specific processing choices retained for MARS and RFDL are
summarised in Table~\ref{tab:rfdl_native_processing}.

\begin{table}[H]
\centering
\caption{Model-specific processing retained for each neural-network method.}
\label{tab:rfdl_native_processing}
\begin{tabular}{lll}
\hline
Component & MARS & RFDL baseline \\
\hline
Source patch size & \(512\times512\) & \(512\times512\) \\
Native network input & \(512\times512\) & \(64\times64\) \\
Source representation & Normalised tanh domain & Positive raw-like amplitude \\
Augmentation morphology & Matched & Matched \\
Post-augmentation mapping & Not required & Raw-like amplitude mapping \\
Learning procedure & Direct mask prediction & CDAE followed by residual detection \\
Signal-preservation loss & Included & Not included \\
\hline
\end{tabular}
\end{table}

This setup controls the source-data split and augmentation distribution while preserving each model's native input preprocessing and learning objective. The comparison therefore primarily reflects differences in model architecture, preprocessing, and signal-preservation strategy rather than differences in the underlying training data or augmentation distribution.

\section{Optional hysteresis mask completion}
\label{app:hysteresis}

The neural detector produces a probability map \(P=\sigma(Z)\), where \(Z\) denotes the output logits. Without hysteresis, the probability map is converted to a binary RFI mask using the threshold \(\tau=0.5\). The pipeline also supports an optional hysteresis mask-completion step intended to connect fragmented RFI structures before replacement.

Let \(\tau_{\rm seed}\) denote the high-confidence seed threshold and let \(\tau_{\rm support}\leq\tau_{\rm seed}\) denote the lower support threshold. The corresponding masks are
\[
S=\mathbb{I}[P>\tau_{\rm seed}],
\qquad
U=\mathbb{I}[P>\tau_{\rm support}].
\]
Here, \(S\) contains high-confidence RFI pixels, while \(U\) contains lower-confidence pixels that may support mask growth.

Starting from \(S\), a fixed number of anisotropic dilation steps is applied, with growth constrained by \(U\). An optional row-support gate prevents growth along frequency rows containing insufficient support. Temporal closing and a final dilation may then be applied to bridge small gaps and extend the completed structure. The original high-confidence seed pixels are always retained.

To prevent excessive mask growth, the implementation limits both the final masked fraction and the fraction of pixels added beyond the seed mask. If either limit is exceeded, the completed mask for that patch is replaced by the original high-confidence seed mask. This guard reduces the risk that hysteresis removes an implausibly large fraction of a patch.

Hysteresis can complete broken narrowband ridges and weak edges around broadband bursts, but it may also increase the number of replaced samples. It is therefore disabled in the main synthetic RFI, FRB-protection, full-filterbank, and real-GMRT results reported in this work. Any diagnostic result obtained with hysteresis is labelled explicitly as HYS-enabled.

\section{SIGPROC synthetic binary-pulsar configuration}
\label{app:sigproc_fake_config}

The synthetic full-filterbank PRESTO benchmark uses the SIGPROC \texttt{fake} program to generate binary-pulsar signals. The spin period, pulse duty cycle, single-pulse peak S/N, dispersion measure, and binary-system parameters are sampled from fixed distributions. The random seed is sampled uniformly from the integer range \([0,2147483647]\). Spin period, pulse width, peak S/N, DM, orbital period, and component masses are sampled log-uniformly, while orbital phase and eccentricity are sampled uniformly.

The parameter \texttt{snrpeak} denotes the signal-to-noise ratio of an individual pulse peak. The generated filterbanks have an output channel spacing of
\[
f_{\rm off}=-0.292968752\,{\rm MHz},
\]
because SIGPROC enforces descending frequency order. The default settings of one summed intensity stream (\(\texttt{nifs}=1\)), a start time of MJD 50000, and dispersion and sampling-time smearing enabled are retained. The complete parameter ranges and fixed generation settings are listed in Table~\ref{tab:sigproc_fake_config}.

\begin{table*}[!t]
\centering
\caption{SIGPROC \texttt{fake} parameters used for the synthetic binary-pulsar full-filterbank benchmark. The pulse-width parameter is expressed as a percentage of the spin period. The listed \(\lvert f_{\rm off}\rvert\) value is supplied to SIGPROC as a positive magnitude; \texttt{fake} converts it to a negative value in the output header so that the first channel is the highest-frequency channel.}
\label{tab:sigproc_fake_config}
\begin{tabular}{lccccc}
\hline
Parameter & 128 \(\mu{\rm s}\) & 256 \(\mu{\rm s}\) & 512 \(\mu{\rm s}\) & 1024 \(\mu{\rm s}\) & 1310 \(\mu{\rm s}\) \\
\hline
\texttt{seed} & 0--2147483647 & 0--2147483647 & 0--2147483647 & 0--2147483647 & 0--2147483647 \\
\texttt{period} (ms) & 1--10 & 1--10 & 1.5--10 & 2.5--10 & 3--10 \\
\texttt{width} (\% of period) & 5--50 & 5--50 & 5--50 & 5--50 & 5--50 \\
\texttt{snrpeak} & 0.08--0.10 & 0.08--0.10 & 0.08--0.10 & 0.08--0.10 & 0.08--0.10 \\
\texttt{dm} (\({\rm pc\,cm^{-3}}\)) & 50--3000 & 50--3000 & 50--3000 & 50--3000 & 50--3000 \\
\texttt{nbits} & 8 & 8 & 8 & 8 & 8 \\
\texttt{nchans} & 4096 & 4096 & 4096 & 4096 & 4096 \\
\texttt{tsamp} (\(\mu{\rm s}\)) & 128 & 256 & 512 & 1024 & 1310 \\
\texttt{tobs} (s) & 300 & 300 & 300 & 300 & 300 \\
\texttt{fch1} (MHz) & 1550 & 1550 & 1550 & 1550 & 1550 \\
Supplied \(\lvert\texttt{foff}\rvert\) (MHz) & 0.292968752 & 0.292968752 & 0.292968752 & 0.292968752 & 0.292968752 \\
\texttt{binary} & Enabled & Enabled & Enabled & Enabled & Enabled \\
\texttt{bper} (h) & 1--10 & 1--10 & 1--10 & 1--10 & 1--10 \\
\texttt{bphase} & 0--1 & 0--1 & 0--1 & 0--1 & 0--1 \\
\texttt{bpmass} (\(M_\odot\)) & 0.5--5 & 0.5--5 & 0.5--5 & 0.5--5 & 0.5--5 \\
\texttt{bcmass} (\(M_\odot\)) & 0.5--5 & 0.5--5 & 0.5--5 & 0.5--5 & 0.5--5 \\
\texttt{becc} & 0--1 & 0--1 & 0--1 & 0--1 & 0--1 \\
\hline
\end{tabular}
\end{table*}

\section{SPECTRALib synthetic-data generation}
\label{app:spectralib_generation}

SPECTRALib is a lightweight Python library for manipulating radio-astronomy filterbank-like data. Of particular relevance to this work, the library provides parametric models for injecting synthetic RFI into filterbank data. We use SPECTRALib version 0.0.23 to construct controlled RFI-contaminated data with known injection parameters.

The relevant public function is \texttt{spectralib.rfi.generate\_rfi}, which operates on a two-dimensional NumPy array with shape \((N_{\rm chan},N_{\rm sample})\). Each call adds a constant amplitude to a selected frequency--time support and clips the resulting data to the 8-bit range:
\[
D'_{c,t}
=
\operatorname{clip}_{[0,255]}
\left[
D_{c,t}
+
A\,\mathbb{I}\bigl((c,t)\in\Omega\bigr)
\right],
\]
where \(A\) is the injected amplitude and \(\Omega\) is the selected frequency--time region.

The function exposes the affected channel width, active duration, amplitude, persistence, repetition count, and duty cycle as parameters. A component whose width equals the total number of channels is broadband, while a smaller width produces narrowband RFI at a randomly selected frequency position. Non-persistent components are placed at a random time, persistent components span the observation, and repeating components alternate between active and inactive intervals. For an active duration \(L_{\rm on}\) and percentage duty cycle \(d\), the inactive duration is
\[
L_{\rm off}
=
\operatorname{round}
\left[
L_{\rm on}\frac{100-d}{d}
\right].
\]

These controls are used to generate five component families: persistent narrowband, repeating narrowband, impulsive narrowband, repeating broadband, and impulsive broadband RFI. Mixed RFI is produced by applying several independently parameterised components to the same data array.

For the patch-level RFI benchmark, SPECTRALib components are injected into clean \(512\times512\) frequency--time patches. The benchmark contains the five individual families and the mixed-RFI family at sampling times of 128, 256, 512, 1024, and \(1310\,\mu{\rm s}\). Five hundred fixed-seed examples are generated for every family and sampling-time combination, giving 15,000 test patches. Durations specified in physical time are converted to sample bins separately at each sampling interval. The identical contaminated patches are used to compare MARS and RFDL.

The public \texttt{generate\_rfi} function returns the contaminated array but not an RFI mask. The benchmark wrapper therefore records the support of each injected component and constructs the ground-truth mask from the union of those supports. This provides exact pixel-level labels independently of background intensity and overlap between components.

For the FRB-plus-RFI protection benchmark, several SPECTRALib components are combined to produce mixed RFI. Component placement is constrained to reduce, but not eliminate, physical overlap with the injected dispersed signal. Signal retention is measured only over FRB support that is not overlapped by injected RFI, while RFI-only recall is measured over injected RFI support that does not overlap the FRB. The benchmark contains 9360 fixed-seed patches spanning the sampling-time, DM, S/N, and pulse-width grid described in the main text.

For the full-filterbank experiment, clean binary-pulsar filterbanks are generated using SIGPROC and SPECTRALib is then used as the contamination layer. Two hundred fixed-seed filterbanks are generated at each of the five sampling times, giving 1000 contaminated inputs. Each contaminated file is supplied unchanged to both the GPU RFI mitigation pipeline and \texttt{filtool}, ensuring that the two mitigation methods are evaluated using identical pulsar, noise, and RFI realisations.

The online training augmentation uses similar RFI morphology classes but is implemented separately in the neural network's normalised input domain. It should therefore be described as SPECTRALib-style augmentation rather than as a direct invocation of \texttt{generate\_rfi}. Finally, because the public SPECTRALib RFI model is based on additive axis-aligned regions, these synthetic tests provide controlled and reproducible ground truth but do not represent every drifting, curved, stochastic, or instrument-specific morphology present in real observations.

\clearpage
\bibliography{sample701}{}

@article{fridman2001rfi,
  title={RFI mitigation methods in radio astronomy},
  author={Fridman, PA and Baan, WA},
  journal={Astronomy \& Astrophysics},
  volume={378},
  number={1},
  pages={327--344},
  year={2001},
  publisher={EDP Sciences}
}

@article{offringa2010post,
  title={Post-correlation radio frequency interference classification methods},
  author={Offringa, AR and De Bruyn, AG and Biehl, M and Zaroubi, S and Bernardi, G and Pandey, VN},
  journal={Monthly Notices of the Royal Astronomical Society},
  volume={405},
  number={1},
  pages={155--167},
  year={2010},
  publisher={Blackwell Publishing Ltd Oxford, UK}
}

@article{petroff2019fast,
  title={Fast radio bursts},
  author={Petroff, Emily and Hessels, JWT and Lorimer, DR},
  journal={The Astronomy and Astrophysics Review},
  volume={27},
  number={1},
  pages={4},
  year={2019},
  publisher={Springer}
}

@article{morello2022iqrm,
  title={IQRM: real-time adaptive RFI masking for radio transient and pulsar searches},
  author={Morello, V and Rajwade, KM and Stappers, BW},
  journal={Monthly Notices of the Royal Astronomical Society},
  volume={510},
  number={1},
  pages={1393--1403},
  year={2022},
  publisher={Oxford University Press}
}

@article{ouyang2024hierarchical,
  title={Hierarchical vision transformers for RFI mitigation in radio astronomy},
  author={Ouyang, Xiaowei and Dreuning, Henk and Mesarcik, Michael and van Nieuwpoort, Rob V},
  journal={BOOK OF},
  pages={24},
  year={2024}
}

@article{van2024remove,
  title={Remove First Detect Later: a counter-intuitive approach for detecting radio frequency interference in radio sky imagery},
  author={van Zyl, Daniel J and Grobler, Trienko L},
  journal={Monthly Notices of the Royal Astronomical Society},
  volume={530},
  number={2},
  pages={1907--1920},
  year={2024},
  publisher={Oxford University Press}
}

@article{mesarcik2022learning,
  title={Learning to detect radio frequency interference in radio astronomy without seeing it},
  author={Mesarcik, Michael and Boonstra, Albert-Jan and Ranguelova, Elena and van Nieuwpoort, Rob V},
  journal={Monthly Notices of the Royal Astronomical Society},
  volume={516},
  number={4},
  pages={5367--5378},
  year={2022},
  publisher={Oxford University Press}
}

@article{offringa2010aoflagger,
  title={AOFlagger: RFI software},
  author={Offringa, AR},
  journal={Astrophysics Source Code Library},
  pages={ascl--1010},
  year={2010}
}

@article{du2024comparison,
  title={A comparison framework for deep learning RFI detection algorithms},
  author={Du Toit, Charl D and Grobler, Trienko L and Ludick, Danie J},
  journal={Monthly Notices of the Royal Astronomical Society},
  volume={530},
  number={1},
  pages={613--629},
  year={2024},
  publisher={Oxford University Press}
}

@article{gu2024radio,
  title={Radio frequency interference detection using efficient multiscale convolutional attention UNet},
  author={Gu, Fei and Hao, Longfei and Liang, Bo and Feng, Song and Wei, Shoulin and Dai, Wei and Xu, Yonghua and Li, Zhixuan and Dao, Yihang},
  journal={Monthly Notices of the Royal Astronomical Society},
  volume={529},
  number={4},
  pages={4719--4727},
  year={2024},
  publisher={Oxford University Press}
}

@article{men2023pulsarx,
  title={PulsarX: a new pulsar searching package-I. A high performance folding program for pulsar surveys},
  author={Men, Yunpeng and Barr, Ewan and Clark, Colin J and Carli, Emma and Desvignes, Gregory},
  journal={Astronomy \& Astrophysics},
  volume={679},
  pages={A20},
  year={2023},
  publisher={EDP Sciences}
}

@article{ransom2002fourier,
  title={Fourier techniques for very long astrophysical time-series analysis},
  author={Ransom, Scott M and Eikenberry, Stephen S and Middleditch, John},
  journal={The Astronomical Journal},
  volume={124},
  number={3},
  pages={1788--1809},
  year={2002}
}

@inproceedings{ronneberger2015u,
  title={U-net: Convolutional networks for biomedical image segmentation},
  author={Ronneberger, Olaf and Fischer, Philipp and Brox, Thomas},
  booktitle={International Conference on Medical image computing and computer-assisted intervention},
  pages={234--241},
  year={2015},
  organization={Springer}
}

@article{lorimer2011sigproc,
  title={SIGPROC: pulsar signal processing programs},
  author={Lorimer, DR},
  journal={Astrophysics Source Code Library},
  pages={ascl--1107},
  year={2011}
}

@article{pritchard2025spiking,
  title={Spiking neural networks for radio frequency interference detection in radio astronomy},
  author={Pritchard, Nicholas J and Wicenec, Andreas and Bennamoun, Mohammed and Dodson, Richard},
  journal={Communications Physics},
  year={2025},
  publisher={Nature Publishing Group UK London}
}

@misc{nvidia_tensorrt_docs,
  author       = {{NVIDIA Corporation}},
  title        = {{NVIDIA TensorRT Documentation}},
  year         = {2026},
  howpublished = {\url{https://docs.nvidia.com/deeplearning/tensorrt/latest/index.html}},
  note         = {Accessed: 2026-07-01}
}

@article{cordes2003searches,
  title={Searches for fast radio transients},
  author={Cordes, JM and McLaughlin, Maura A},
  journal={The Astrophysical Journal},
  volume={596},
  number={2},
  pages={1142--1154},
  year={2003}
}

@article{offringa2012morphological,
  title={A morphological algorithm for improving radio-frequency interference detection},
  author={Offringa, AR and Van De Gronde, JJ and Roerdink, JBTM},
  journal={Astronomy \& astrophysics},
  volume={539},
  pages={A95},
  year={2012},
  publisher={EDP Sciences}
}

@article{yuan2022categorize,
  title={Categorize radio interference using component and temporal analysis},
  author={Yuan, Mao and Zhu, Weiwei and Zhang, Haiyan and Huang, Shijie and Xue, Mengyao and Li, Di and Yue, Youling and Wang, Pei and Niu, Jiarui and Hu, Yuxuan and others},
  journal={Monthly Notices of the Royal Astronomical Society},
  volume={513},
  number={4},
  pages={4787--4801},
  year={2022},
  publisher={Oxford University Press}
}

@article{carels2019development,
  title={Development of production-ready GPU data processing pipeline software for AstroAccelerate},
  author={Carels, Cees and Ad{\'a}mek, Karel and Novotn{\`y}, Jan and Armour, Wesley},
  journal={arXiv preprint arXiv:1912.07704},
  year={2019}
}

@misc{sigpyproc_software,
  author       = {{sigpyproc developers}},
  title        = {sigpyproc: Pulsar and FRB Data Analysis Library},
  year         = {2026},
  howpublished = {GitHub software repository},
  url          = {https://sigpyproc3.readthedocs.io/en/latest/index.html},
  version      = {stable}
}

@misc{spectralib_software,
  author       = {White, Jack},
  title        = {{SPECTRALib}: Synthetic Pulsar Emission, Contamination,
                  and Transients Radio Astronomy Library},
  year         = {2026},
  howpublished = {GitHub software repository},
  url          = {https://github.com/jack-white1/SPECTRALib}
}

@misc{tensorflow2015-whitepaper,
title={ {TensorFlow}: Large-Scale Machine Learning on Heterogeneous Systems},
url={https://www.tensorflow.org/},
note={Software available from tensorflow.org},
author={
    Mart\'{i}n~Abadi and
    Ashish~Agarwal and
    Paul~Barham and
    Eugene~Brevdo and
    Zhifeng~Chen and
    Craig~Citro and
    Greg~S.~Corrado and
    Andy~Davis and
    Jeffrey~Dean and
    Matthieu~Devin and
    Sanjay~Ghemawat and
    Ian~Goodfellow and
    Andrew~Harp and
    Geoffrey~Irving and
    Michael~Isard and
    Yangqing Jia and
    Rafal~Jozefowicz and
    Lukasz~Kaiser and
    Manjunath~Kudlur and
    Josh~Levenberg and
    Dandelion~Man\'{e} and
    Rajat~Monga and
    Sherry~Moore and
    Derek~Murray and
    Chris~Olah and
    Mike~Schuster and
    Jonathon~Shlens and
    Benoit~Steiner and
    Ilya~Sutskever and
    Kunal~Talwar and
    Paul~Tucker and
    Vincent~Vanhoucke and
    Vijay~Vasudevan and
    Fernanda~Vi\'{e}gas and
    Oriol~Vinyals and
    Pete~Warden and
    Martin~Wattenberg and
    Martin~Wicke and
    Yuan~Yu and
    Xiaoqiang~Zheng},
  year={2015},
}

@article{paszke2019pytorch,
  title={Pytorch: An imperative style, high-performance deep learning library},
  author={Paszke, Adam and Gross, Sam and Massa, Francisco and Lerer, Adam and Bradbury, James and Chanan, Gregory and Killeen, Trevor and Lin, Zeming and Gimelshein, Natalia and Antiga, Luca and others},
  journal={Advances in neural information processing systems},
  volume={32},
  year={2019}
}
\bibliographystyle{aasjournalv7}



\end{document}